\documentclass[a4paper,12pt,fleqn,usenatbib,usAMS]{mnras}

\usepackage{hyperref}
\usepackage[dvips]{graphicx}
\usepackage[latin1]{inputenc}
\usepackage{guit}
\usepackage{color}
\usepackage{times}
\usepackage{natbib}
\usepackage{setspace}
\usepackage{amsmath}
\usepackage{amsfonts}
\usepackage{amssymb}
\usepackage{multirow}
\usepackage{multicol}
\usepackage{aas}
\usepackage{booktabs}
\usepackage{adjustbox}
\usepackage{float}
\usepackage{graphics}

\usepackage{bm}		% Bold maths symbols, including upright Greek
\usepackage{pdflscape}	% Landscape pages
\usepackage[T1]{fontenc}
\usepackage{ae,aecompl}
\usepackage{subfigure,lscape,color}
\newif\ifAMStwofonts
\AMStwofontstrue
\definecolor{red}{rgb}{1,0.,0.}
\usepackage{float}
\title[MOS spectroscopy of High-z galaxies]{MOS spectroscopy of protocluster candidate galaxies at z= 6.5}

\author[R. Calvi et al.]{R. Calvi$^{1,2}$\thanks{E-mail: rcalvi@iac.es, rosa.calvi@gmail.com}, J.M. Rodr\'iguez Espinosa$^{1,2}$\thanks{E-mail:jre@iac.es}, J.M. Mas-Hesse$^3$,   
\newauthor  K. Chanchaiworawit$^{4,5}$, R. Guzman$^{4,6}$,  E. Salvador-Sol\'e$^6$, J. Gallego$^7$, 
\newauthor A. Herrero$^{1,2}$, A. Manrique$^6$,  and A. Mar\'in Franch$^8$ \\
$^1$Instituto de Astrof\'isica de Canarias, E-38205 La Laguna, Spain \\ 
$^2$Depto. de Astrof\'isica, Universidad de La Laguna, E-38206 La Laguna, Spain\\
$^3$Centro de Astrobiolog\'ia - Depto. de Astof\'isica (CSIC-INTA), Madrid, Spain \\
$^4$Department of Astronomy, University of Florida, 211 Bryant Space Science Center, P.O. Box 112055, Gainesville, FL, 32611-2055, USA\\
$^5$National Astronomical Research Institute of Thailand, 260  Moo 4, T. Donkaew, A. Maerim, Chiangmai, 50180, Thailand\\
$^6$Institut de Ciencies del Cosmos, Universitat de Barcelona, UB-IEEC. Mart\'i Franqu\'es 1, E-08028 Barcelona, Spain\\ 
$^7$Depto. de Astrof\'isica y CC de la Atm\'osfera, Universidad Complutense de Madrid, Spain\\ 
$^8$CEFCA, Plaza san Juan 1, E-44001 Teruel, Spain 
}
\date{Accepted 2019 July 30, Received 2019 July 22; in original form 2019 February 15}

\pubyear{2019}

\begin{document}

\label{firstpage}
\pagerange{\pageref{firstpage}--\pageref{lastpage}}
\maketitle

\begin{abstract}
The epoch corresponding to a redshift of z $\sim 6.5$ is close to full re-ionisation of the Universe, and early enough to provide an intriguing environment to observe the early stage of large-scale structure formation. It is also en epoch that can be used to verify the abundance of a large population of low luminosity star-forming galaxies, that are deemed  responsible for cosmic re-ionisation. Here, we present the results of follow-up multi-object spectroscopy using OSIRIS at Gran Telescopio Canarias (GTC) of 16 Ly$\alpha$ emitter (LAE) candidates discovered in the Subaru/XMM Newton Deep Survey. We have securely confirmed 10 LAEs with sufficient signal-to-noise ratio of the Ly$\alpha$ emission line. The inferred star formation rates of the confirmed LAEs are on the low side, within the range 0.9-4.7 M$_{\odot}$ yr$^{-1}$. However, they show relatively high Ly$\alpha$ rest frame equivalent widths. Finally we have shown that the mechanical energy released by the star formation episodes in these galaxies is enough to create holes in the neutral hydrogen medium such that Lyman continuum photons can escape to the intergalactic medium, thus contributing to the re-ionisation of the Universe. 

\end{abstract}


\section{Introduction}

The last decades have seen a strong interest for studies of the early Universe. Many people have shown that using characteristic spectral signatures, such as the Lyman-$\alpha$ (Ly$\alpha$) line of Ly$\alpha$ emitters (LAEs) and the prominent rest-frame UV continuum of Lyman-Break galaxies (LBGs), is effective in searching for distant large-scale structures and protoclusters beyond $z\geq 3$ (e.g., \citet{Naidu2018,Kakiichi2018}). Additionally, there is increasing evidence that low luminosity star-forming galaxies in the early Universe were the main culprits for the re-ionisation of the Universe \citep{Ouchi09,Bouwens10,Robertson15}, which was mostly completed by $z\sim 6$ (e.g.,\citet{Fan06,Inoue18}).

It has been proposed that the best place to detect these UV-dominated sources in the Epoch of Re-ionisation is around bright quasi-stellar objects (QSO),  Radio galaxies, or  intense starburst galaxies. These sources are signposts of enhanced matter concentration or, possibly, a galaxy cluster in the process of formation--a protocluster (e.g., \citet{{venemans2004},{debreuck2002},{lefevre1996},{carilli1997},{pentericci1997},{overzier2006b}}). 

This Paper is part of an observational program aimed to (1) explore the LAEs population near the end of the Epoch of re-ionisation; and (2) search for the clustering properties of these high-z galaxies. We have observed the field around two luminous $z\sim6.5$ star-forming LAEs (separated by $\sim$300 pkpc) discovered by \citet{Ouchi10} in the Subaru/XMM Newton Deep Survey (SXDS). We have performed deep photometric and spectroscopic observations of the region containing these two LAEs, using the instrument called OSIRIS at the Gran Telescopio Canarias (GTC).  Paper I \citep{Kritt17} gives a complete description of the photometric observations, including strategy, data reduction, sample selection, interlopers filtering and statistical analysis. This paper (Paper II) deals with the follow-up multi-object spectroscopy of a subset of the LAE candidates reported in Paper I.  A third paper \citep{Chanchaiworawit2019}, discusses the clustering properties of these LAEs, the mass of the overdensity, and its growth via linear spherical collapse. In particular, we have shown that the sources discussed in this paper belong to a proto-cluster \citep{Chanchaiworawit2019}. In \citet{Chanchaiworawit2019} we have discussed 1) the presence of a virialised core, and 2) the evolution of the whole proto-cluster, which will virialise by z$\sim 0.84$, and by z= 0 will reach a mass comparable to that of the Coma cluster \citep{Chanchaiworawit2019}. The size of the proto-cluster, encompassing these sources, is $10873\pm 1679$\,cMpc$^3$. 

While this paper was being prepared, \citet{silverrush7} as part of the SILVERRUSH survey, did present a large photometric survey and spectroscopic follow up including our field as a subregion of the SXDS field. That work presented spectroscopic results of a number of sources in common with our LAE candidates. The redshift measurements of the common sources are consistent within $1\sigma$ uncertainty, as we will show later.

This paper presents the results of two observing runs of multi-object spectroscopy (MOS) with OSIRIS/GTC. Sections \ref{sec:obs} and \ref{sec:redux} describe the observations and the data reduction and how we dealt with the noise in these deep and low signal to noise observations. Section 4 describes the detected sources and their properties. Finally, Section 5 discusses the main results. We adopt the $\Lambda$CDM concordant Universe model ($\Omega_\Lambda$ = 0.7, $\Omega_{M}$=0.3, and h = 0.7). Magnitudes are given in the AB system \citep{Oke83}.

%%%%%%%%%%%%%%%%%%%%%%%%%%%%%%%%%%%%%%%%%%%%%%%%%%%%%%%%%%%%%%%%%%%%%%%%%%%%%%%%%%%%%%%%%%%%

\section{Observations} \label{sec:obs}

The photometric observations took advantage of 3 medium-band filters from the SHARDS program  \citep{pablo2013}, namely the  F883(33.6), F913(27.8), and F941(33.3) filters.  The filter names give information on the central wavelength of each filter in nanometer. The values in parenthesis indicate the width of each filter also in nanometer. The redshifted (z= 6.5) Ly$\alpha$ emission line would appear as a detection in F913, while it would be  non-detected in the F883 filter nor in any bands blueward of F883. In order to reach an adequate level of sensitivity allowing the study of the number density and clustering of such a faint population, we observed through the filters, F883, F913, and F941, with total exposure times in each band of 12.25, 10.78, and 11.30 hours, respectively. The exposure times per frame in F883, F913, and F941 were 350s, 400s and 300s, respectively. We used a 6-point dithering pattern, tracing a parallelogram with 8 arcsec base and 16 arcsec height. The dithering pattern was used in order to obtain complete information of the fluctuation pattern and strength of the sky background. The photometric observations were carried out during semesters 2011B and 2012B on the GTC, with OSIRIS in its imaging mode. The median seeing at $\sim 9000\AA$ and the median air-mass during the observing runs were 0.7'' and 1.20, respectively. Our survey resulted in 45 LAE candidate sources selected photometrically in the SXDS-North field. We categorised them into 2 classes based on their F913 flux profiles: class-I candidates would exhibit flux profiles resembling compact galaxies in the F913 filter, with the peak of the profile near the centre, and almost circular in shape; class-II LAE candidates show signs of deviations from the ideals of class-I, such as having noise contamination, or peak flux skewed from the centre. Extensive details of the photometric observations and data processing are given in \citet{Kritt17}.

\subsection{Mask Design}

\begin{figure}
\centering
 \includegraphics[width=\columnwidth]{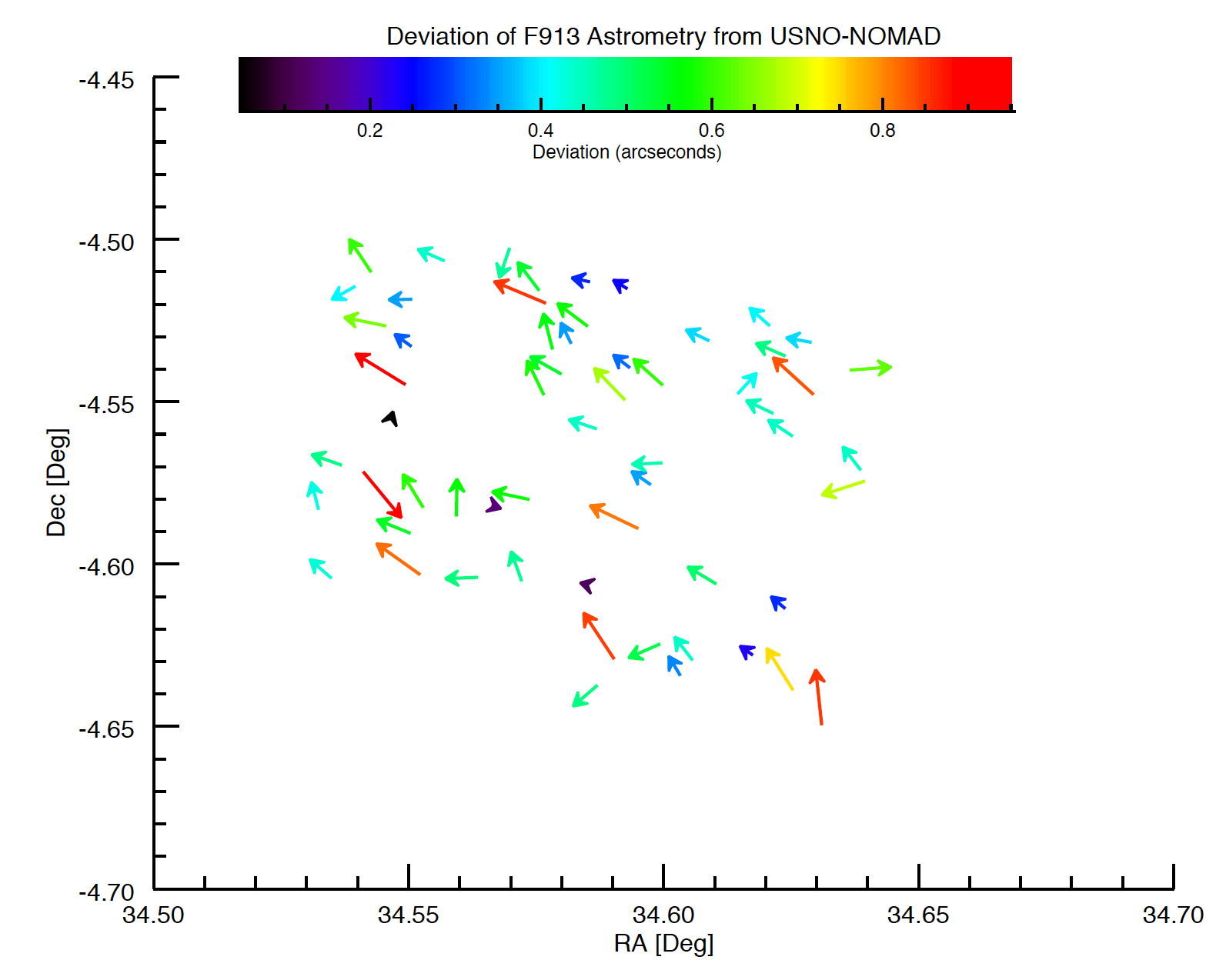}
 \caption{Differences in astrometry between the F913 image and the USNO-NOMAD catalogue \citep{usnob}. The colours of vectors represent the differences in astrometry in units of arcseconds. The vectors point to the positions of USNO catalogue's stars with F913's positions as references.}
 \label{usno}
\end{figure}

 We used the following strategies to select the best set of LAE candidates, guiding stars, and filler objects for the spectroscopic follow-up. First, we performed a reliability test of our astrometry during the photometric selection. By cross-checking the stars' Right Ascension (RA) and Declination (Dec) against the values reported in the USNO catalogue \citep{usnob}, we assessed both systematic or random shifts in astrometry.  The systematic shift between the two catalogues was $\rm \Delta_{sys} = 0.5''$, while the random shift was negligible ($\rm \Delta_{rand} < 0.1''$) as shown in Figure~\ref{usno}. The correction to systematic shift was achievable by aligning multiple guiding stars within the OSIRIS-MOS mask. With the very small random shift ($\rm \Delta_{rand} << seeing$), we are confident that slits on the OSIRIS-MOS mask can correctly target the faint LAE candidates and filler  objects.

Out of the 45 LAE candidates photometrically selected from the SXDS-North field \citep{Kritt17}, we chose 16 objects to fill one OSIRIS-MOS mask. We included the LAE candidates that fit within the FOV of $\rm 7.5'\times4'$ to guarantee full spectral coverage between $\lambda = 8500-9500 \mathring{A}$. The selection of the sources was made to achieve spatial uniformity and a good distribution across all the F913 magnitude bins. However, to optimise the success rate of the program, we gave higher priority to class-I LAE candidates first. Then, we filled the available spaces of the MOS mask with class-II LAE candidates. We also used 6 guiding stars with z'-band AB magnitude between 18-19 magnitude. The MOS mask included 1 already confirmed luminous LAE from \citet{Ouchi10}, ID NB921-N-79144, or C1-01 in our nomenclature, and 3 filler objects. In the end, the MOS mask consisted of 26 objects in total (1 known LAE (Ouchi) + 16 LAE candidates + 6 guiding stars + 3 fillers), without spectral superpositions from any pair of slits. The 2''-aperture circular holes were used for the guiding stars and for centring the mask with high precision. The science object slits are rectangular with 1'' width, which is about the typical seeing of spectroscopic nights at the GTC, and 15'' in length to optimise spectral resolution and facilitate the quality of sky subtraction. We designed the mask to optimise the number of LAE candidates seen in a single pointing. A single pointing was preferred to allot all of the observing time to just one mask.

\begin{figure}
\centering
 \includegraphics[width=\columnwidth]{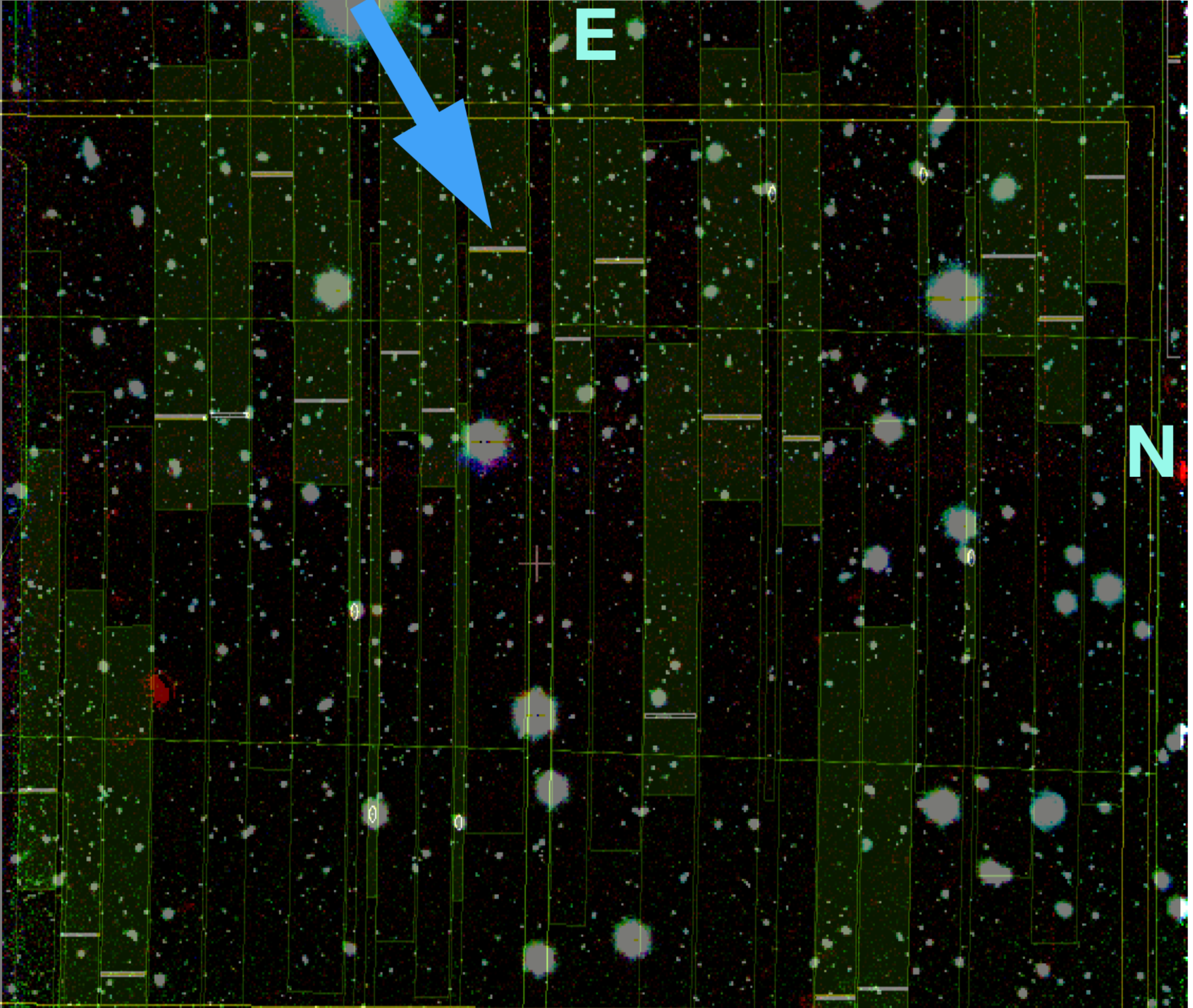}
 \caption{The mask used for the MOS in the OSIRIS/GTC observations. The mask includes 20 science objects and 6 guiding stars. The blue arrow indicates the slit associated with C1-01, one of the confirmed LAEs from \citet{Ouchi10}, included for comparison.}
 \label{fig1}
\end{figure}

\subsection{MOS Observations}\label{mos_obs}

The observations were carried out in service mode (queue mode) in two separate semesters (2016B and 2017B; see Table~\ref{Obs}) at the GTC with OSIRIS, the GTC's optical imager and spectrograph, \citep{Cepa03} in its MOS (Multi-Object Spectroscopy) mode. We used the R2500I grism, which produces a spectral resolution of $2.7 \mathring{A}$ at $\lambda = 9000 \mathring{A}$. The OSIRIS field of view (FOV) is a mosaic of two CCDs with a small gap in between.  The FOV is $7.5'\times 6.0'$ in MOS mode, with a plate scale of $0.127''$/pix. To get the full spectra into the CCD chips the sources should be located within $\rm 7.5'\times4'$. With the 2$\times$2 binning that we adopted, the plate scale became $0.254''/pix$. The design of the mask was done with a special software tool - the OSIRIS Mask Designer Tool (MD). For the mask design, we aimed to optimise the number of objects in the available FOV. The mask consisted of 20 rectangular slits, 1''-wide and at least 15''-long (up to 25''-long) for LAE candidates and fillers. There were also 6 circular holes of 2'' in diameter for up to six stars, used for aligning the mask on the sky. We did not use any dithering pattern along the slit. The resulting spectra cover a spectral range between 6500-9700 $\mathring{A}$.  Figure \ref{fig1} shows part of the MOS mask used in our program with the slits oriented in the North-South direction (i.e.,  a position angle of -90 deg on the sky).

\begin{table}
	\begin{tabular}{@{}lccccc}
    \hline
    \hline    
	Obs. Date & R.A. & Dec. & Grism & Int. Time & Seeing \\
	                 & [hh:mm:ss] & [dd:mm:ss] &   &  [sec] & \\	
    \hline	
    2016.10-12 & 02:18:18.4 & -04:34:45.0 & R2500I & 90000 & $\leq 1"$ \\
    2017.09    & 02:18:18.4 & -04:34:45.0 & R2500I & 43200  & $\leq 1"$\\   
     \hline   
    \end{tabular}
\caption{Details of the OSIRIS-MOS observation runs.  The coordinates refer to the center of the field. The observations consisted of 36 individual observing blocks (OBs),  of one hour each.  Each OB consisted of three science frames with an exposure of 1150 seconds each. The rest was dedicated to overheads and calibrations. The observations were performed in the second half of 2016 and 2017.}
\label{Obs}
\end{table}

\section{Data Reduction} \label{sec:redux}

The spectra were processed using IRAF routines \citep{iraf1}. The process included de-bias, flat fielding, cosmic rays removal, extraction of spectra from individual slits, wavelength calibration, and sky subtraction. Given the low signal-to-noise (S/N) and complex nature of the observed spectra, we decided to perform the data reduction independently at both participating institutions --the Instituto de Astrof\'{i}sica de Canarias (IAC) and the Department of Astronomy, University of Florida (UF).

\subsection{The IAC data reduction} 

The number of Observing Blocks (OBs) obtained from the first run (2016) was 26. However, 6 of them presented important defects, and specifically the Ly$\alpha$ line of the C1-01 source (Ouchi's source) was not visible. So for further analysis, we only used the 20 remaining OBs. We reduced and calibrated the frames with IRAF routines \citep{iraf1}. As the sources were expected to be quite faint, we tried our best to not introduce additional noise at this stage of the data reduction. Each OB, consisting of three frames, was first median averaged. The bias subtraction and flat-fielding had been performed before for each OB separately. Cosmic rays were removed by median combining the three frames. We used a reference flat for each OSIRIS' CCD to normalise differential gains and correct geometrical distortions. Finally, we removed the vertical distortions in both CCDs using 2 reference coordinate maps, then we extracted the individual spectra from the corrected images. The wavelength calibration process was done individually for each of the 17 LAE slits (3 slits were fillers), using a list of sky lines. As the Far-Optical to NIR sky is full of OH emission lines the wavelength calibration was performed by matching the observed sky emission features with the optical-NIR sky emission lines from the LRIS catalogue compiled by Sergey Zharikov of Universidad Nacional Aut\'onoma de Mexico (UNAM) \footnote{http://www.astrossp.unam.mx/~resast/standards/NightSky/skylines.html}. 

After the two-dimensional (2D) spectra were wavelength calibrated with 17 or 18 sky lines, we performed the sky subtraction. First, utilising the IRAF's BACKGROUND task, we fitted and subtracted the spectra with a cubic-spline function of $\rm 4^{th}$ order as well as a Chebyshev polynomial of $\rm 4^{th}$ order along the spatial direction. During the fitting process, we excluded a small region equivalent to the typical seeing, where the LAE should be. This method omitted the flux of the faint object from the background, thus avoiding removing it along with the sky. If an emission feature could be visually spotted, we extracted a one dimensional (1D) spectrum by summing the flux along a 5-pixel window ($\rm 1.25''$). If residuals from the first sky subtraction were visible, we smoothed the spectra using the IRAF's BOXCAR task. This task smooths out the images using a rectangular-filter of fixed dimensions (usually $\rm 3\times 3 \: pix^2$, rarely $\rm 4\times 4 \: pix^2$). 

There were 11 OBs in the observing run of 2017B. We did not find any defective OBs during the second semester of the spectroscopic follow-up. The data reduction and calibration processes were performed in a similar fashion. Since the frames from 2016B and 2017B were slightly shifted from each other, they were analysed separately. Only after the individual spectra were extracted they were combined to maximise the signal-to-noise ratio (S/N). To accurately identify the detections of LAEs, along the spectral direction, we adopted the following guidelines: (1) the detections should be located between $\rm 9000\mathring{A} \leq \lambda \leq  9280 \mathring{A}$ (the width of the F913 filter); (2) they should not exhibit any continuum or emission blue-ward of the Ly$\alpha$ line; (3) there should not be prominent sky lines near the potential lines. 

\begin{figure}
\centering
\subfigure[C1-01]{
 \includegraphics[width=\columnwidth]{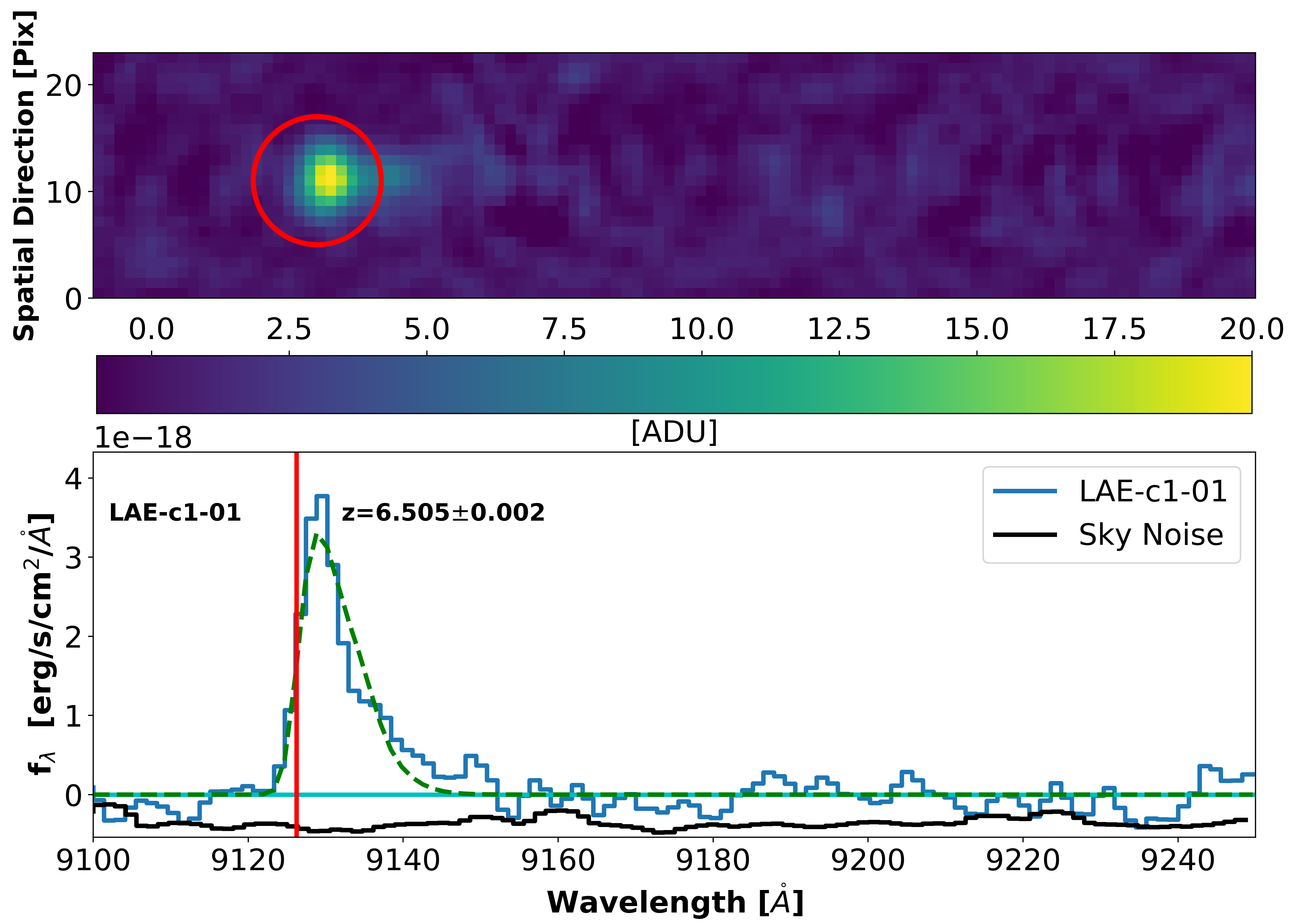}}
 \caption{Two- and one-dimensional spectra of LAE-c1-01. The blue line is the collapsed 1D spectrum. The vertical red solid line indicates the measured spectroscopic redshift. The corresponding redshift value and its uncertainty are written next to the red solid line. The black solid line is the 1D sky background, scaled down so it does not overlap with the LAE's 1D spectrum.}
 \label{Ouchi}
\end{figure}

The spectrum of the OUCHi source (C1-01 in our code) was readily spotted. That spectrum is shown in Fig.~\ref{Ouchi}. Note that the redshift is somewhat different, because our computation of the skewness, i.e., we have not used the peak of the Ly$\alpha$ line to derive the redshift as will be explained below. The spectrum is nonetheless much better than the one in \citet{Ouchi10}, thus the fluxes and luminosities are better determined here. 

\subsubsection{The {\it NoiseChisel} Approach}

The visual inspection is a subjective process. Due to the faintness of the LAE candidates, relying solely on this approach may not be the most reliable and robust way to declare a detection. Thus, we used the {\it NoiseChisel} code \citep{Mohammad15} to assess the reliability of our identifications and to guide our eyes in selecting possible identifications. The code implements both ``noise-based'' and ``noise-parametric'' techniques for detecting faint objects, including very low-surface brightness regions of a bright object buried under the noise. A detailed description of the code can be found in \citet{Mohammad15}. Briefly, the basic idea behind this algorithm is that it assumes that contiguous pixels of signal, beyond what is expected from cosmic rays, are indications of actual sources. Therefore, the code check for large patches of contiguous pixels, while keeping unchanged the noise fluctuations. {\it NoiseChisel} requires only a few constraints and does not rely on any regression analysis. This ability allows faint clumps of data with $\rm S/N\leq 4$ to be accurately identified (with 99\% confidence level). {\it NoiseChisel} mask the sky-lines. However, it does not mask the high-frequency background noise or residuals. Given that the sky lines were always at the same wavelength and only their intensity differs between different slits, we built a unique mask for all frames and just changed the ratio height/length of the regions that need to be masked for each slit (as long as the centroids were accurately defined). This step allowed us to adjust the width of the mask to match the sky lines in each frame.

We only considered regions mostly free from sky-lines. So we masked out the prominent sky lines and trimmed the pixels at the edges of the images. Then, we were left with a clean image for the {\it NoiseChisel} code to detect low S/N emission clumps. The optional values to run the code are listed in the Table\ref{param}. We considered for  most of them the default parameters of GNU Astronomy Utilities 0.3. We just modified the following parameters: tilesize, openingngb, dthresh, detsnminarea, detquant, dilate, and interpnumngb.

\begin{table*}
\centering
\begin{tabular}{lccl}
  \hline
  \hline
 \multirow{4}{*}{Input} & hdu  & 1 & Extension name or number of input data.\\
 & khdu & 1 & HDU containing Kernel image. \\
 & minskyfrac & 0.7 & Min. fraction of undetected area in tile. \\
 & minnumfalse & 100 & Minimum number for S/N estimation.\\
  \hline
  \multirow{7}{*}{Tessellation (tile grid)} &  tilesize & 25,25 & Regular tile size on dim.s (FITS order).\\
& numchannels & 1,1 & No. of channels in dim.s (FITS order).\\
  & remainderfrac & 0.1 & Fraction of remainder to split last tile. \\
  & workoverch & 0 & Work (not tile) over channel edges.\\
  & interponlyblank & 0 & Only interpolate over the blank tiles.\\
  & interpnumngb & 1 & No. of neighbors to use for interpolation.\\
  & largetilesize & 200,200 & Sim. to --tilesize, but for larger tiles.\\
  \hline
  \multirow{1}{*}{Output} & tableformat & txt & Formats: 'txt', 'fits-ascii', 'fits-binary'\\
  \hline
  \multirow{2}{*}{Operating modes} & minmapsize & 1000000000 & Minimum bytes in array to not use ram RAM.\\
  & continueaftercheck & 0 & Continue processing after checks.\\
  \hline
  \multirow{14}{*}{Detection} & mirrordist & 1.5  & Max. dist. (error multip.) to find mode.\\
  & modmedqdiff & 0.01 & Max. mode and median quant diff. per tile.\\
  & qthresh & 0.3 &  Quantile threshold on convolved image.\\
  & smoothwidth & 3 &  Flat kernel width to smooth interpolated.\\
  & erode & 2 & Number of erosions after thresholding.\\
  & erodengb & 4 & 4 or 8 connectivity in erosion. \\
  & noerodequant & 0.9331 & Quantile for no erosion. \\
  & opening & 1 & Depth of opening after erosion. \\
  & openingngb & 4 & 4 or 8 connectivity in opening. \\
  & sigmaclip & 3,0.2 & Sigma multiple and, tolerance or number. \\
  & dthresh & -0.1 & Sigma threshold for Pseudo-detections.\\
  & detsnminarea & 5 & Min. pseudo-detection area for S/N dist.\\
  & detquant & 0.89 & Quantile in pseudo-det. to define true.\\
  & dilate & 1 & Number of times to dilate true detections.\\
  \hline
  \multirow{6}{*}{Segmentation} & segsnminarea & 15 & Minimum area of clumps for S/N estimation.\\
  & segquant & 0.95 & S/N Quantile of true sky clumps.\\
  & keepmaxnearriver & 0 & Keep clumps with peak touching a river.\\
  & gthresh & 0.5 & Multiple of STD to stop growing clumps.\\
  & minriverlength & 15 & Minimum len of useful grown clump rivers.\\
  & objbordersn & 1 & Min. S/N for grown clumps as one object.\\
  \hline
  \end{tabular}
\caption{Optional values of the NoiseChisel code as applied to our set of spectra.\label{param}}
\end{table*}

\begin{figure}%
\centering
\subfigure[Sky mask on the noise image]{%
\label{fig:first1}%
\includegraphics[scale=0.4]{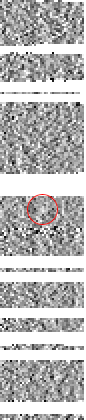}}%
\qquad\qquad
\subfigure[Clumps detection]{%
\label{fig:first2}%
\includegraphics[scale=0.4]{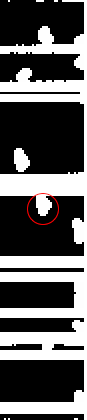}}%
\caption{Example of {\it Noise Chisel} detection process on LAE-c1-15. {\bf Left:} a part of the 2D spectrum with the sky emission lines masked. {\bf Right:} the detected clumps with S/N larger than the preset threshold. The grey-scale shows higher ADU counts in lighter region. The red circles indicate the detected source.}
\end{figure}

The code, then, assesses the probability for each contiguous clump of pixels of being real. An example of a 2D spectrum with masked sky lines and the final spectrum showing the detected clumps of pixels is illustrated in Figure \ref{fig:first1} and \ref{fig:first2}, respectively. Attending to the most probable clumps, we selected the clumps that matched the expected position of the LAE candidates on 2D spectra and checked for the signature of Ly$\alpha$ emission. Therefore, the final selection for a positive detection of a LAE were made by combining the results of {\it NoiseChisel} and visual inspection.

\subsection{The UF's data reduction}\label{uf_redux}

As already mentioned in Section \ref{mos_obs}, the proposed OSIRIS-MOS observations were carried out on two different semesters (2016B-2017B) due to the extended integration time required to achieve optimal S/N for all objects. The reduction processes of the spectroscopic frames were done in 2 stages- (1) CCD processing and (2) spectroscopic calibration. We started the process by stitching 2 OSIRIS CCDs together for all science and calibration frames, using the GTCMOS pipeline (see also \citep{gtcmos1}) developed and maintain by Divakara Mayya of the Instituto Nacional de Astrof\'{i}sica, \'{O}ptica y Electr\'{o}nica (INAOE). We obtained a master bias image from median stacking. Using the master bias image, all science and flat frames were bias-subtracted (de-biased). The flat frames were median combined and normalised to obtain a  master normalised flat image. Then, we flat-fielded the science frames, cleaned out bad pixels and cosmic rays using the IRAF's COSMICRAYS task, and sliced them into 2D strips corresponding to the individual slits. 

At this point, we started the spectroscopic calibration process by performing the wavelength calibration for each slit. We decided that the automated script of the GTCMOS pipeline may not be sufficient for our purpose of detecting very faint emission lines. Thus, we performed the following spectroscopic calibration using IRAF routines on the individual slit frames. The slit frames were aligned and median combined to obtain the master slit image. We used the plentiful OH sky emission lines ranging from 7400 to 10500 $\mathring{A}$ in the master slit image to perform wavelength calibration. The wavelengths of prominent sky emission lines were also obtained from the LRIS catalogue of NIR sky emission lines. The sky emission lines used in the calibration process are between 25-40 lines per frame. We fitted the dispersion function of each slit image by a Chebyshev polynomial of $4^{th}$ order and applied the fitted solution to rectify all frames of the individual slits. 

Due to the expected low S/N of the science targets, we took an extra cautions in sky subtraction processes by introducing iterative sky subtraction methods. We performed the first iteration of sky subtraction on the rectified 2D spectra by smoothing the frames with a $3\:pix\times 3\:pix$ box and fitting Chebyshev polynomial of $4^{th}$ order along the spatial direction with 2$\sigma$-clipping. We combined the first sky-subtracted frames to form the master science image for each slit. We looked for any detection within the same position along the slit, within the 9000 to 9280 $\mathring{A}$ range (corresponding to the width of F913 band), and sufficiently far away from sky emission lines. We recorded the xy-positions of the potential detections for all slits. Then, the second iteration of sky subtraction used the Chebyshev polynomial of $6^{th}$ order (or higher depending on the length of the slits--15''-20'') to fit and obtain the background by omitting $\pm$3 pixels around the recorded positions of the initial detections. 

Next, we performed the sky subtraction by shifting the position of the initial detections by $\pm$1 pixel to find the best solution that maximise the S/N of the detection, while minimising residuals of sky emissions.The best version of sky-subtracted frames of each slit were average combined with 3$\sigma$ clipping. Then, we collapsed the reduced 2D spectrum of each object into 1D spectrum. We relied on an iterative process of fine tuning to obtain the best possible quality of the LAEs' 1D spectra which optimise S/N and minimise sky contamination in both spectral and spatial directions. Considering the observed physical diameter of LAEs of up to $\rm \sim 1''$ and the typical seeing of spectroscopic nights, which was $\rm \leq 1''$, we used an extraction window of $\rm \sim1.5''$ to sum up fluxes along the spatial direction of all pixels within this window. The size of the extraction window allowed us to optimise the total flux counts and minimise possible sky contamination. The process started with locating the position along the slit of the centroid of each emission line. Then, the extraction window was defined as $\rm \pm 3\:pix$ along the y-axis from the centroid. So, the total flux of a LAE or LAE candidate is the sum along the y-axis of this extracted spectrum. Since not all science frames were useful, we discarded all the science slit frames in which we could not detect the bright LAE from \citet{Ouchi10}. This was likely due to drifting and pointing problems of those particular frames. In the end, we combined a total of 61 science frames (from the 20 OBs in 2016B and 11 OBs in 2017B) to obtain the highest S/N 2D spectra for the LAE and LAE candidates at z=6.5 of our program.

\subsection{Spectral Quality Assessment}  \label{compare}

\begin{figure}
\centering
 \includegraphics[width=\columnwidth]{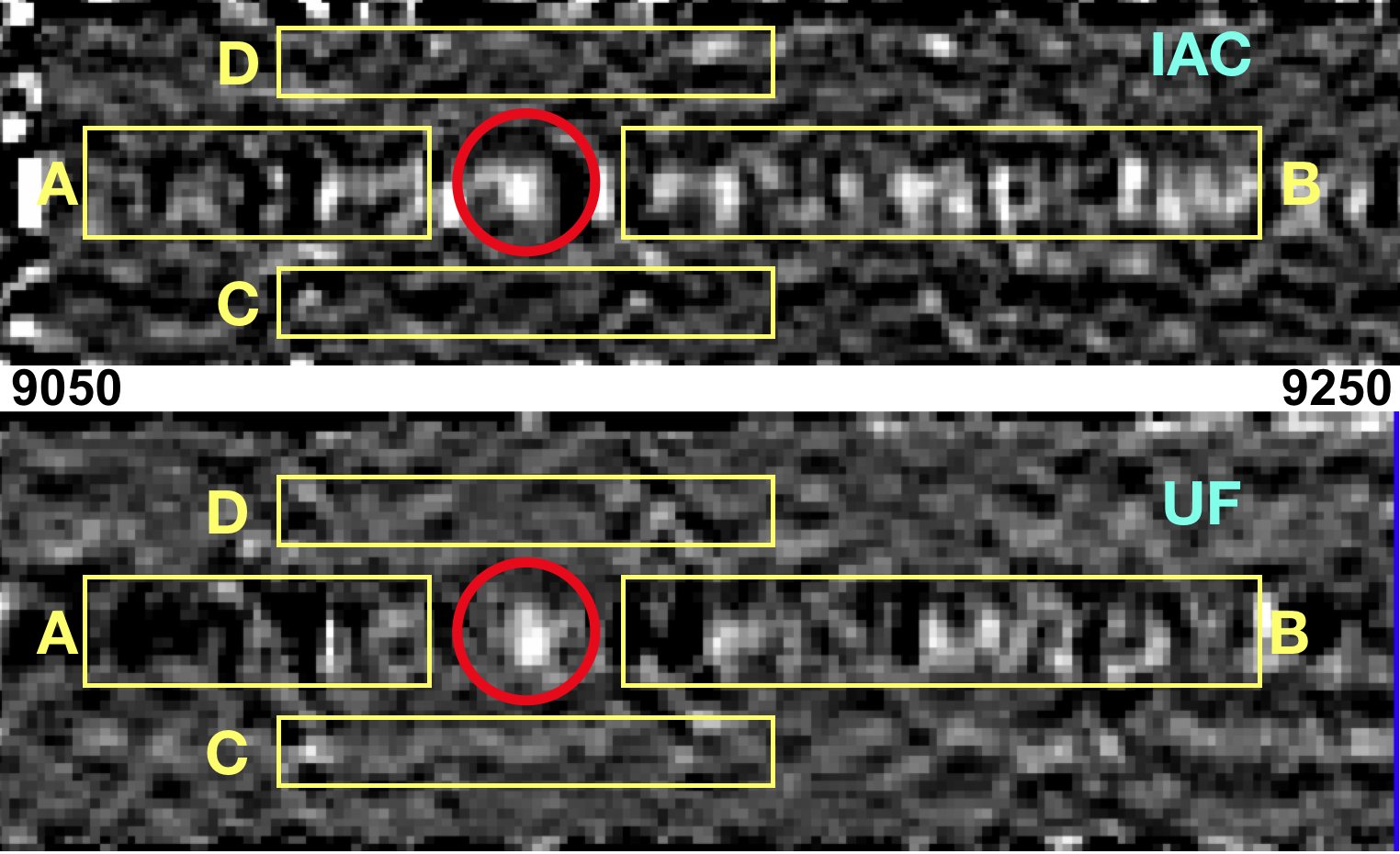}
 \caption{An example of the quality of the final reduced spectra between the IAC's and UF's reduction processes. The upper and lower panels are the 2D spectra of the detected LAE-C1-15 from the IAC's and UF's sides, respectively. The red circle shows the detected Ly$\alpha$ emissions. The yellow rectangles labeled A and B represent the region blue- and red-ward of the Ly$\alpha$ line along the spectral direction, respectively. While the yellow rectangles labeled C and D represent the region adjacent to the LAE along the spatial direction.}
 \label{c115_compare}
\end{figure}

After the two institutions performed their own data reduction independently, we compared the quality of the final reduced spectra on the basis of the S/N of the detections, and sky residuals. An example of the comparison and quality assessment process are illustrated in Figure~\ref{c115_compare}, using one of the newly confirmed LAEs, namely LAE-C1-15, as representative. First, we compare the sky residuals by focusing on the rectangular boxes labeled A and B as shown in Figure~\ref{c115_compare}. The UF's reduced 2D spectra demonstrated suppressed sky residual along the spectral direction (both blue- and red-ward of the Ly$\alpha$ emission features) as compared to the IAC's counterparts. The result suggested the effectiveness of iterative sky subtraction processes. While the leftover sky residuals exhibited by the IAC's 2D spectra suggested that the non-iterative sky subtraction might be too aggressive or underestimating the sky levels under the LAEs, which were omitted during the fitting process. The sky residuals might be the indirect results of the wavelength calibration with insufficient number of sky lines as well. 

Next, we focus on the RMS noise along the spatial direction as indicated in the C and D boxes. The noise levels in these regions dictate the S/N of the detection. We found no significant difference between noise levels in these regions between the IAC's and UF's spectra. Thus, we concluded that the S/N of Ly$\alpha$ emission features from both institution are similar. However, note that some of the IAC's 2D spectra show signs of `ghost' or negative trench in either C or D box. This, again, is indicative of an overestimated background and likely an aggressive sky subtraction.

\subsection{Flux Calibration} \label{flux_cal}

The last step before we could analyse the spectroscopic results was to put the observed spectra into a physical unit. A spectroscopic standard, G191-B2B, was observed in both GTC semesters with extremely high S/N. We reduced the spectroscopic standard (henceforth, the standard) frames in the same manner as the science frames, except the need for iterative sky subtractions. However, since the standard was taken with the long-slit mode and a wider slit width ($2.5"$), different flat frames corresponding to the standard were used. The flat-fielded, de-biased standard frames were then cut to be only 400 pixels wide and 2020 pixels long (100'' $\times$ 2800 $\mathring{A}$). This reduced size allowed for quicker wavelength calibration, while maintaining the spectral range of interest and optimising the sky subtraction. The wavelength calibration on the standard frames was performed similarly to the science frames. However, the broader slit width produced lower spectral resolution. Identifying centroids of sky emission lines suffered larger uncertainties. Nevertheless, the much larger spatial width allowed for an adequate efficiency for wavelength calibration. The sky subtraction process was done with a single iteration, since we could conveniently locate the position of the standard star on the frames.
    
Considering the brightness of G191-B2B, we used $2''$ extraction window, sufficient to sum up all the standard's flux into the 1D spectrum.  Knowing the exposure time of the standard, we matched the 1D spectrum of the spectroscopic standard in the unit of ADU \: s$^{-1}$ to the available flux density catalogue of \citet{oke1990}. We obtained the flux transformation relation from $\rm ADU\: s^{-1}$ to $\rm ergs\:s^{-1}\:cm^{-2}\:\mathring{A}^{-1}$. We, thus, achieved the flux calibrated 1D spectra of the LAE candidates, ready for analysis and measurement.
    
Both the IAC's and UF's methods have already proven to be quite robust by demonstrating the measured Ly$\alpha$ flux of LAE-C1-01 in excellent agreement with the photometrically estimated values from previous studies (e.g., \citep{Kritt17,Ouchi10}) as shown in Table \ref{tab_spec}.

\section{Results} \label{results}

This section presents the set of final spectra we consider adequate, which correspond to low luminosity sources undergoing strong star bursts. We have measured their fluxes, star formation rates and luminosities. Finally, using the Ly$\alpha$ escape fractions  computed in \citet{Chanchaiworawit2019}, we have calculated the intrinsic Ly$\alpha$ Luminosities, which we have used to estimate the UV continuum from these sources, using the typical ratios of starburst galaxies. Finally, we calculate the mechanical energy that these sources are releasing into the medium.

\subsection{The final spectra} 
\label{fin_spec}

\begin{figure}
\centering
 \includegraphics[width=\columnwidth]{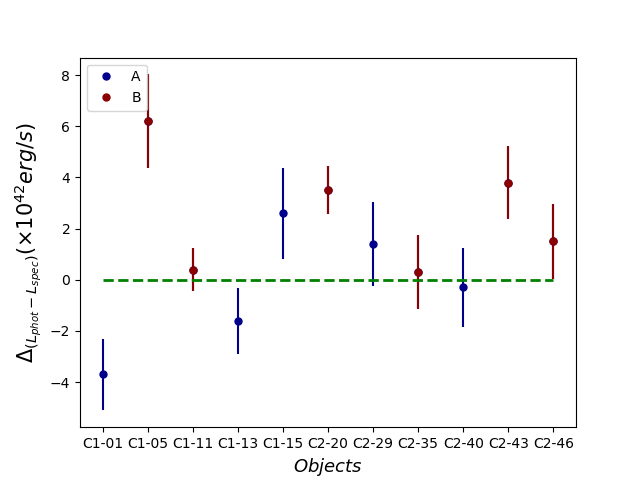}
 \caption{The difference between the photometric, measured in the filter F913, and spectroscopic luminosities  for both A (blue square points) and B (diamond red points) sources. Notice that the Ouchi source (C1-01) and the C1-05 show the largest discrepancies.}
 \label{difference}
\end{figure}

The final collection of sources detected with various degree of confidence from both the IAC's and the UF's catalogues is presented here. Both methods have pros and cons. Thus, we reached the following compromise. We rely on the results of the IAC's {\it {Noise Chisel}} as a guide to identify the location of the Ly$\alpha$ emission features. Then we tried to agree on the quality of the spectra. To that end  we adopted a set of criteria to rank the detected sources into 3 categories--excellent (A), acceptable (B), and marginal (C). The LAE candidates were ranked by their spectroscopic reliability grades (A, B, and C, with A being the most reliable). The grading criteria were based on (1) the shape of their spectra resembling a P-Cygni profile in the 1D spectra; (2) the sizes of Ly$\alpha$ emission features are about $1''$ ($\sim$ seeing of a spectroscopic night), but neither too extended nor too compact in 2D (to discern the real detection from sky and cosmic ray residuals); (3) the locations of emission lines are fairly far away from strong sky emission lines or their wings and; (4) the Ly$\alpha$ luminosity measured spectroscopically must be comparable to the photometric estimate (see Figure~\ref{difference}). If a detection satisfies all of the four criteria, it is given the letter grade ``A''. If a detection fails any one or more criteria, it is given grade ``B'' or ``C'', respectively. 

There are 4 LAEs in grade A, namely C1-13, C1-15, C2-29, and C2-40 as shown in Figure~\ref{gradeA}.  Furthermore, there are 6 LAEs in grade B, namely C1-05, C1-11, C2-20, C2-26, C2-35, C2-43, and C2-46 as shown in Figure \ref{gradeB}. These 10 newly confirmed LAE candidates (grades A and B) are considered spectroscopically confirmed and will serve as the basis for further analysis. Note that some LAEs in grade B and most of LAEs in grade C usually fail to meet some quality criteria, namely the proximity to some sky lines or the lack of consistency between the spectroscopically measured and photometrically estimated fluxes. These results are to be expected for low S/N detections, such as in our case of LAEs at z=6.5. Nevertheless, these criteria do not exclude the possibility that some of grade C LAE candidates can be real LAEs at z=6.5. The decision to move forward with the most reliable 10 LAEs is our choice to be conservative, especially when dealing with such complex and low S/N data. The rest of LAE candidates (grade C sources) also exhibit various degrees of detection as shown, for instance, in Figure \ref{gradeC}. For example, C1-06, C1-07, C2-17 and C2-26  show distinguishable detections on the 2D spectra and signs of the P-Cygni profile on the 1D spectra. However, two of them were on top of the wings of prominent sky emissions and fail to match the photometrically estimated Ly$\alpha$ fluxes, which may be due to the proximity of the sky lines. Hopefully with better resolution they can be confirmed in the future. Overall, the spectroscopic success rate is as expected. We have confirmed 10 out of the 16 LAE candidates, which is a success rate of 62.5\% or non detection rate of 37.2\%. This percentage is slightly above  $\frac{1}{3}$, which is the spuriousness rate previously determined in the photometry \citep{Kritt17}. \\

Note that some LAE candidates are in common with the newly published results from the SILVERRUSH survey \citep{silverrush7}, namely  C1-11, C1-13, C2-26, C2-27, C2-43, C2-46. While C2-47 was observed spectroscopically by SILVERRUSH it was not in our mask. Moreover, C2-26 and C2-27 are in our grade C objects, as they were close to a sky line. The redshift measurements from the 2 programs are in good agreement, within $1\sigma$ uncertainty, considering the different method of measuring the redshift of the emission line (i.e., measure at the peak of the observed Ly$\alpha$ emission or at the centroid of the Gaussian after applying skewness, see Section \ref{lya_flux}). The fact that some of our sources have been spectroscopically detected by other authors gives credibility to the spectra we have obtained. Figure\,~\ref{redshift} shows a histogram with the redshifts (line-of-sight velocities) distribution. Note that there is a relatively smooth distribution albeit skewed towards a  redshift close to $z=6.52$, which is consistent with the expected redshift of the protocluster discussed in \citet{silverrush7} and \citet{Chanchaiworawit2019}.

\begin{table*}
\resizebox{\textwidth}{!}{ 
\begin{tabular}{@{}l c c c c c c c c}   % 9 columns, alignment for each
 \hline
Object   &  R.A. & Dec. &  z$\rm ^{spec}$ & L$\rm (Ly\alpha)^{phot}$ & F$\rm (Ly\alpha)^{spec}$  & Skewness     &    L$\rm (Ly\alpha)^{spec}$   &  Grade \\
 & [$\rm hh:mm:ss$]  &  [$\rm dd:mm:ss$]   &  & [$\rm 10^{42}\:ergs \:\: s^{-1}$]    &  [$\rm 10^{-17}\:ergs \:\: s^{-1}\: cm^{-2}$]  &  alpha       &    [$\rm 10^{42}\:ergs \:\: s^{-1}$] &    \\  
  (1)  &   (2)  &  (3)   &  (4) & (5) & (6) &  (7) & (8) & (9) \\
        \hline           
LAE-C1-01(NB92-N79144) & 2:18:27.0300 & -4:35:08.267 & 6.508$\pm$0.002 & 11.3$\pm$1.4 & 3.11$\pm 0.62 $ & 4.51$\pm$1.45 &  15.0$\pm$0.05 & A (OUCHI)    \\
LAE-C1-05 & 2:18:08.2186 & -4:38:00.294 & 6.592$\pm$0.002 & 7.4$\pm$1.8  & 0.24$\pm 0.05 $ & 2.37$\pm$2.16  &  1.2$\pm$0.3  &  B  \\
LAE-C1-11 & 2:18:22.4661 & -4:36:54.651 & 6.564$\pm$0.003 & 3.2$\pm$0.6 & 0.57$\pm 0.11 $ & 2.98$\pm$1.58    &    2.8$\pm$0.6   &  B    \\
LAE-C1-13 & 2:18:26.8276 & -4:31:21.075 & 6.548$\pm$0.002 & 3.5$\pm$0.8 & 1.05$\pm 0.21$ & 4.21$\pm$1.43 & 5.1$\pm$1.0 &  A    \\
LAE-C1-15 & 2:18:22.5997 & -4:35:27.820 & 6.512$\pm$0.002 & 4.9$\pm$1.7 & 0.47$\pm 0.09$ & 2.38$\pm$0.56 & 2.3$\pm$0.5   &  A    \\
LAE-C2-20 & 2:18:21.8270 & -4:32:53.379 & 6.499$\pm$0.002  & 4.9$\pm$0.9        & 0.29$\pm 0.06 $ & 3.25$\pm$1.60 &    1.4$\pm$0.3   &  B    \\
LAE-C2-29 & 2:18:25.1010 & -4:31:02.156 & 6.549$\pm$0.017 &  4.7$\pm$1.5  & 0.67$\pm 0.13 $ & 2.19$\pm$0.97   &    3.3$\pm$0.7   &  A     \\
LAE-C2-35 & 2:18:06.7520 & -4:32:22.535 & 6.564$\pm$0.005 & 1.9$\pm$1.2 & 0.35$\pm 0.07 $  &  26.6$\pm$3.2  & 1.6$\pm$0.8 &  B \\
LAE-C2-40 & 2:18:28.9901 & -4:30:45.280 &  6.556$\pm$0.002 & 3.8$\pm$1.3 & 0.83$\pm 0.17 $ & 19.5$\pm$832.9 &  4.1$\pm$0.8   &  A      \\
LAE-C2-43 & 2:18:24.1854 & -4:35:40.977 & 6.604$\pm$0.002 & 4.8$\pm$1.4      & 0.20$\pm 0.08 $ &  6.37$\pm$4.67   &    1.0$\pm$0.2   &  B     \\
LAE-C2-46 & 2:18:29.0698 & -4:36:41.021 & 6.553$\pm$0.002 & 3.9$\pm$1.3      & 0.49$\pm 0.10 $ & 2.66$\pm$1.76 & 2.4$\pm$0.7 & B    \\

\hline
\end{tabular}
}
\caption{List of all confirmed LAEs with their parameters; Columns: (1) name.; (2) \& (3) R.A. and Dec.; (4) spectroscopic redshift; (5) photometrically estimated Ly$\alpha$ luminosity; (6) spectroscopically measured Ly$\alpha$ flux;  (7) Skewness (alpha parameter); (8) spectroscopically measured Ly$\alpha$ luminosity; (9) detection grade: A = secured detection, B = sound detection, C = marginal detection. Note that for Ouchi's source we are adding the name given in \citet{Ouchi10}}.
\label{tab_spec}
\end{table*}

\subsection{Ly$\alpha$ Flux and Redshift Measurements} \label{lya_flux}

Traditionally, a Gaussian distribution or Normal distribution function suffices in fitting an emission line. Nevertheless, one of the critical features of the Ly$\alpha$ emission line at high redshift is the asymmetry of the line profile due to the high optical depth of interloping neutral hydrogen gas. Thus, using a normal symmetric Gaussian profile to fit these emission lines would be inappropriate \citep{{pcygni1},{pcygni2}}. However, making assumptions on the neutral hydrogen (HI) optical depth, the interloping intergalactic medium (IGM), and the intrinsic Ly$\alpha$ luminosity is beyond the scope of this paper. Indeed, any assumption made here would create a bias and introduce more uncertainty to our calculations. Thus, we choose the skew Gaussian profile for the fitting function, since it can be tweaked and twisted to fit arbitrary shapes of the detected Ly$\alpha$ emission lines without relying on any astrophysical assumption. Thus, we fitted the 1D spectra of all spectroscopically detected LAEs with a skew Gaussian distribution function with a form as shown in Equation \ref{eq:fitting_skewg2},
     
\begin{equation}
\centering
f(x') \sim \frac{2}{\sigma} \Theta (\alpha x')N(\sigma^2, x_0)\times H + B,  
\label{eq:fitting_skewg2}
\end{equation}

where, $\Theta (\alpha x')$ and $\rm N(\sigma^2, x_0)$ denote the modifying function with skewness factor, $\rm \alpha$, and the normal distribution parts of the function \citep{skew1,skew2,skew3}. The parameters, $\rm x_0$, $\rm H$, and $\rm B$ in Equation \ref{eq:fitting_skewg2} are the median (before applying skewness) of the normal distribution, scale height (height of the distribution), and base value (in case of non-zero median sky background and/or continuum). The modifying function, which skews the traditional normal distribution to the left or right has the form shown in Equation \ref{eq:fitting_skewg3}, 
      
\begin{equation}
\centering
 \Theta (\alpha x')= \frac{1}{2} [1+ erf(\frac{\alpha (x - x_0)}{\sqrt{2\sigma^2}})].
\label{eq:fitting_skewg3}
\end{equation}

The reason behind choosing this function in fitting the Ly$\alpha$ emission line is not only to avoid making assumption on the astrophysical processes of LAEs at $\rm z=6.5$, but also the fact that the modifying function does a perfect job in mimicking the transmission functions of Ly$\alpha$ photons at high redshifts, especially within the EoR and when $\rm \alpha>>1$. We utilised the $\sc scipy.optimize$ package in $\sc Python$ to fit this skew Gaussian distribution function to the observed 1D spectra of our LAE detections.
      
Now that the best-fitted skew Gaussian functions to all observed Ly$\alpha$ emission lines were available, we obtained the total Ly$\alpha$ flux by simply integrating the best-fitted functions. However, before integrating, we subtracted the base of the best-fitted function to get rid of non-zero sky background or any continuum that might show up in the fitting process. In most cases, however, the base was close to zero. The Ly$\alpha$ luminosity was then computed as expressed in Equation \ref{eq:lum_lya}, 

\begin{equation}
\centering
 L^{obs}_{Ly\alpha} = (4\pi d^2_{L})\times F_{Ly\alpha},
\label{eq:lum_lya}
\end{equation}

where $\rm d_L$ is the luminosity distance at $\rm z=6.5$, which is $\rm 63,326.5 \:Mpc$. We can then calculate fluxes and observed Ly$\alpha$ luminosities of LAEs with this formalism, where the flux uncertainties can be directly assessed from the goodness of fit. In case of the redshift measurements, the best fitted $x_0$ value and its uncertainty give us the centre wavelength of what would be the Ly$\alpha$ emission line before passing through a high optical depth IGM. Thus, we can calculate the redshift of each LAE by simply, $\rm z_{spec} = \frac{x_0}{1216 \mathring{A}} -1$. The measured redshifts, the observed Ly$\alpha$ fluxes, the skewness values and the observed Ly$\alpha$ luminosities (both spectroscopically measured and photometrically estimated) are presented in Table \ref{tab_spec}.

\subsection{Consistency Test of the Final Spectra}

\begin{figure}
\centering
 \includegraphics[width=\columnwidth]{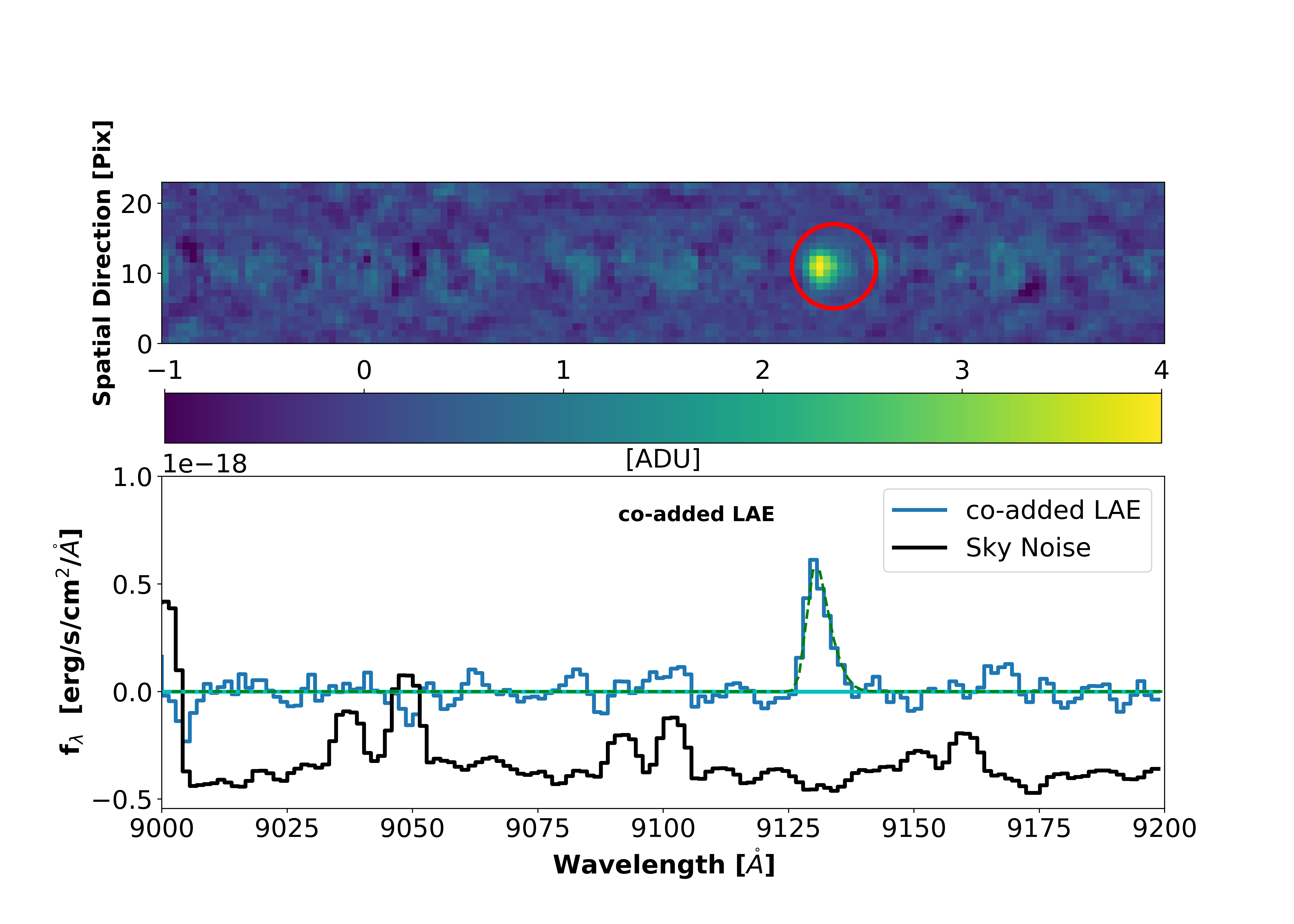}
 \caption{The co-added 2D and 1D spectrum of the 10 LAEs at z=6.5 with reliable detections (LAEs with the letter grade A and B).The blue solid line is the 1D spectrum. While the black solid line is the sky background. The skewness factor is 2.93$\pm1.39$}
 \label{stack}
\end{figure}

\begin{figure}
\centering
 \includegraphics[width=\columnwidth]{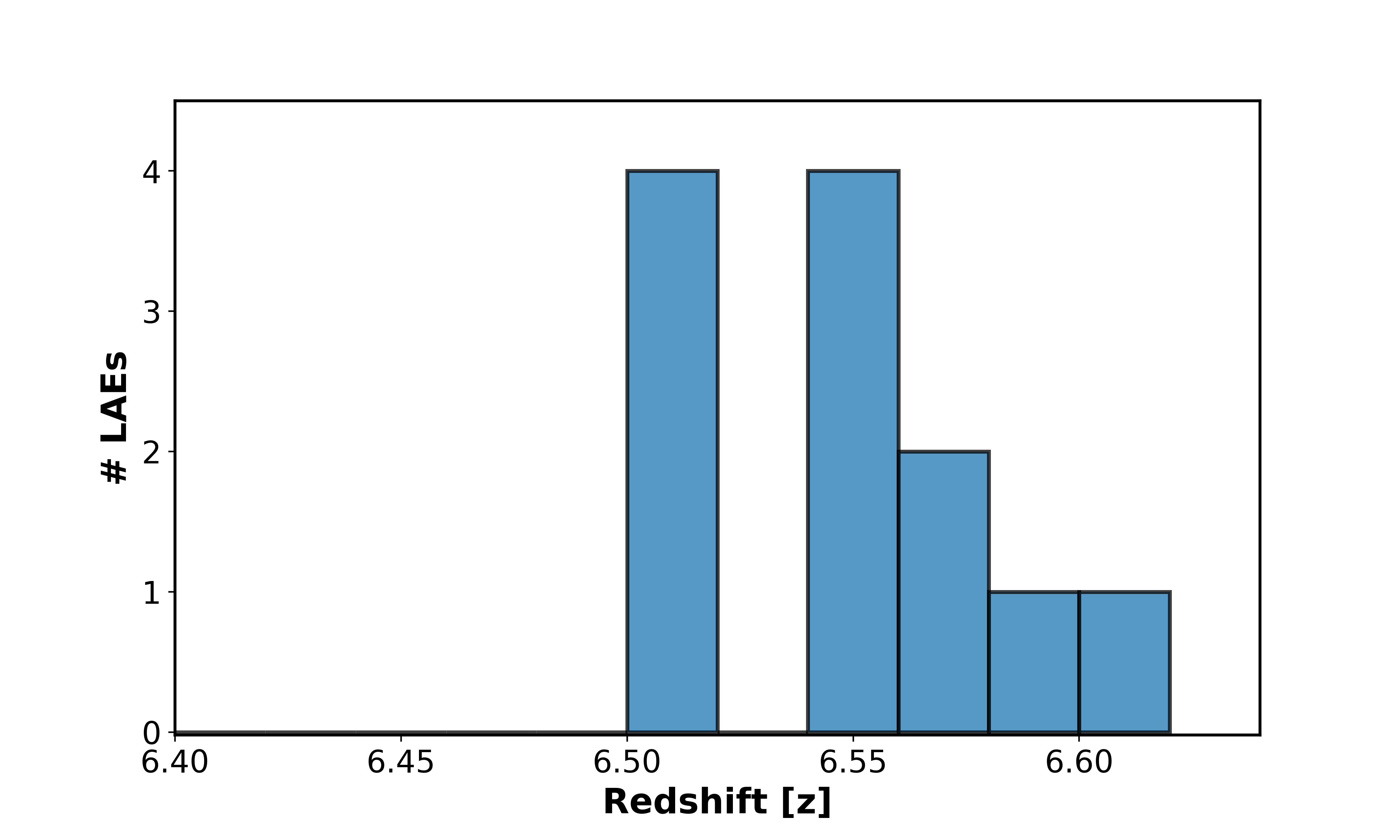}
 \caption{Redshift distribution of the 10 new spectroscopically confirmed LAEs and 2 previously confirmed by \citet{Ouchi10}. The bin size is $\Delta$z=0.02.}
 \label{redshift}
\end{figure}

\begin{figure*}
\centering
\subfigure[C1-13]{
\includegraphics[width=0.4\linewidth]{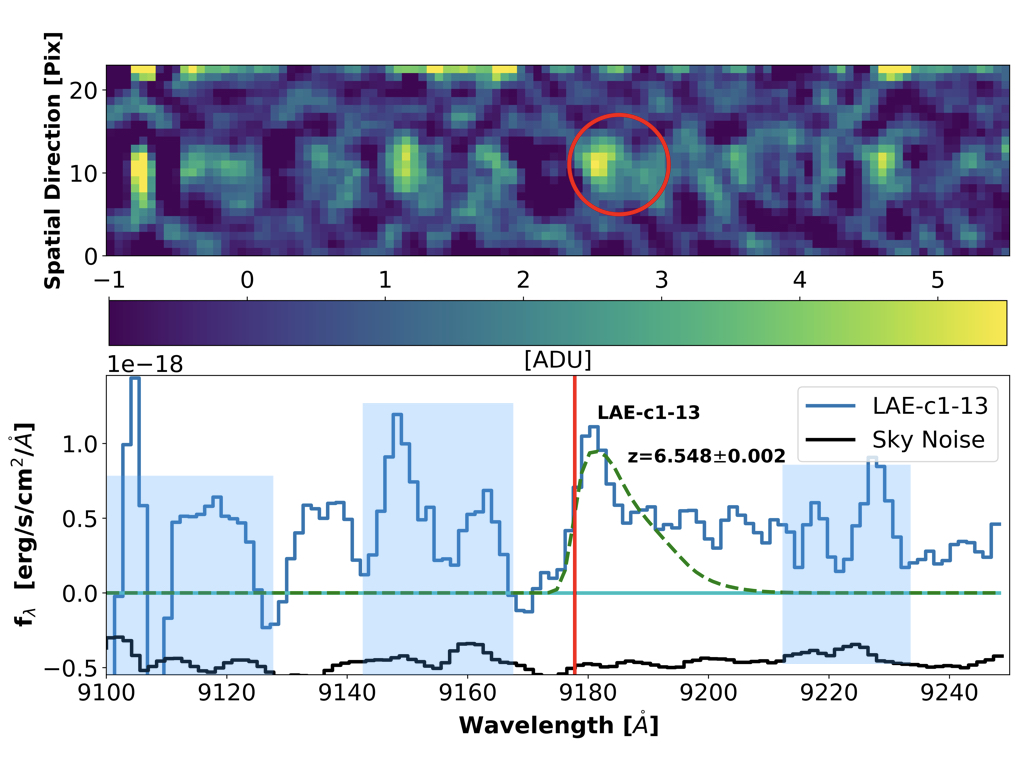}}
\subfigure[C1-15]{
\includegraphics[width=0.4\linewidth]{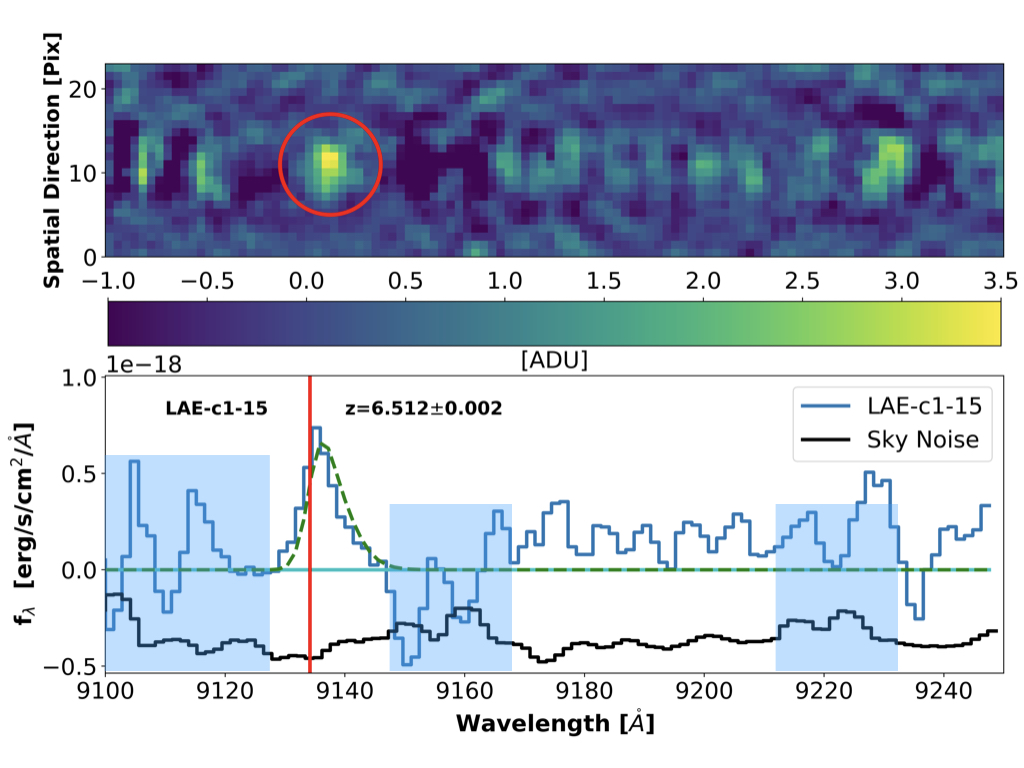}}
\subfigure[C2-29]{
\includegraphics[width=0.4\linewidth]{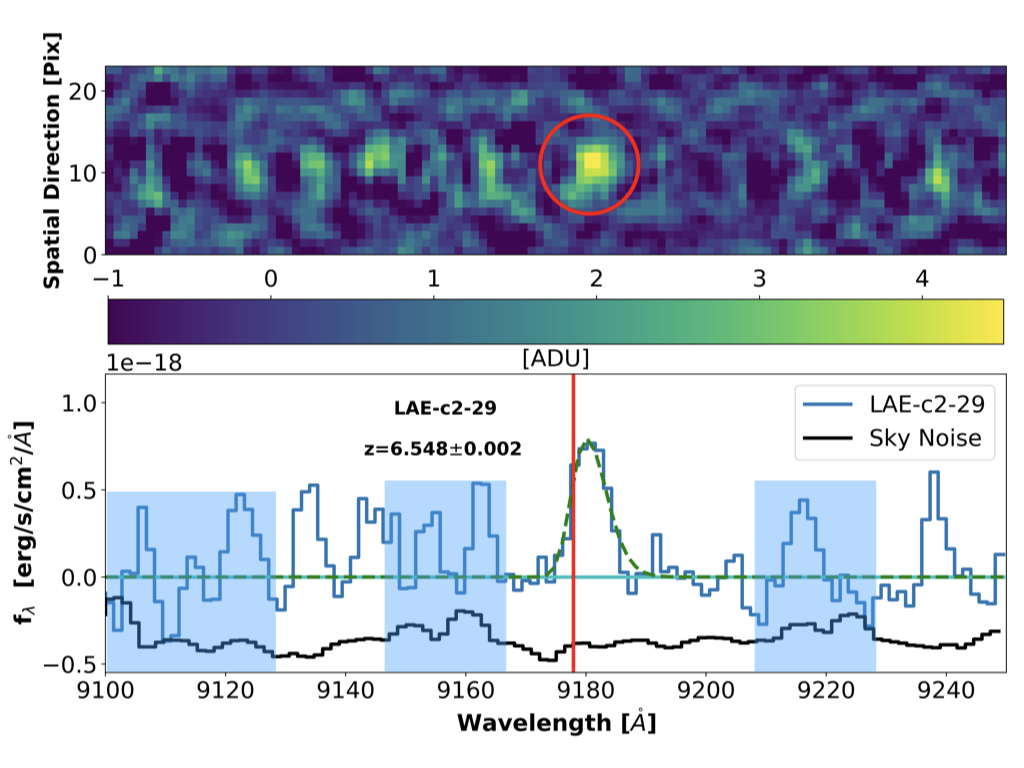}}
\subfigure[C2-40]{
\includegraphics[width=0.4\linewidth]{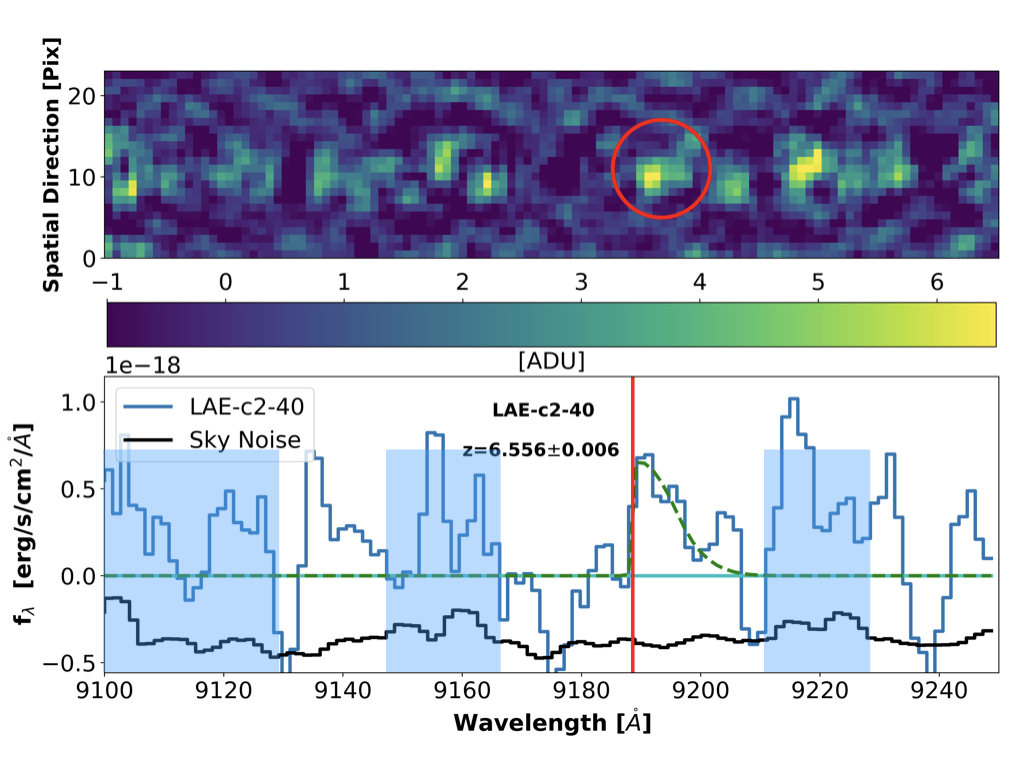}}
\caption{2D and 1D spectra of spectroscopically confirmed LAEs with the letter grade A. The blue solid lines are the extracted 1D spectra. The vertical red solid line indicates the spectroscopic redshifts of each object. The black solid lines are the sky background, measured using the same extraction window as the Ly$\alpha$ emission features, but scaled down to avoid interfering with the object's spectrum. The shaded blue areas indicate those regions where the sky lines are prominent.}
\label{gradeA}
\end{figure*}

\begin{figure*}
\centering
\subfigure[C1-05]{
\includegraphics[width=0.4\linewidth]{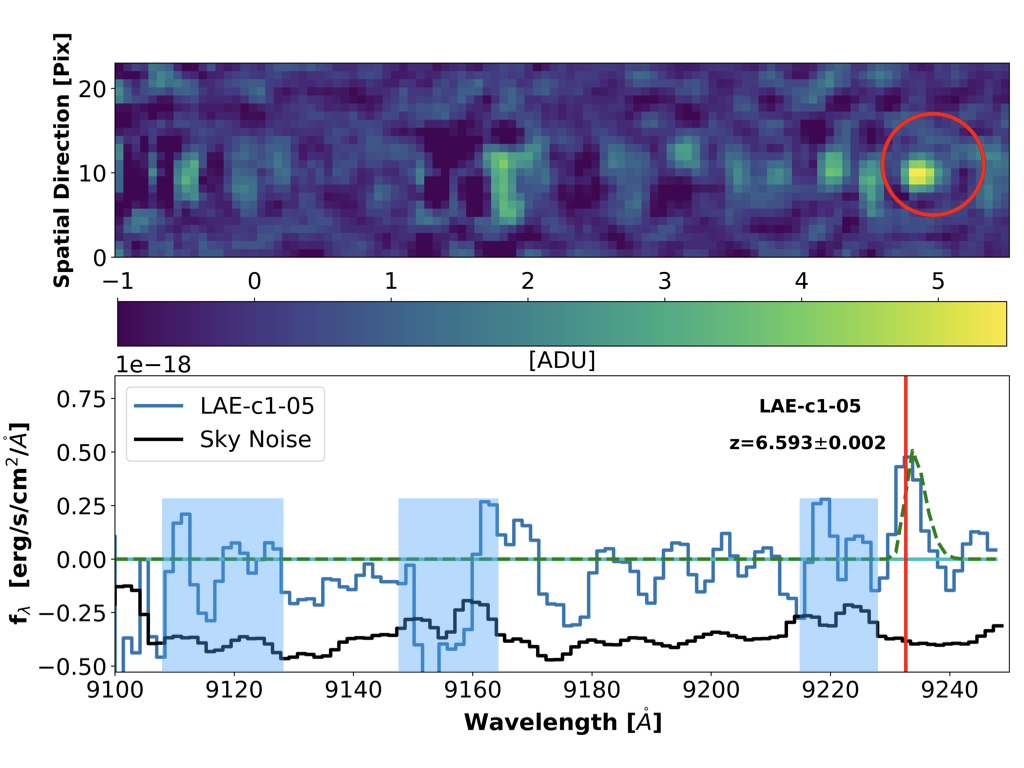}}
\subfigure[C1-11]{
  \includegraphics[width=0.4\linewidth]{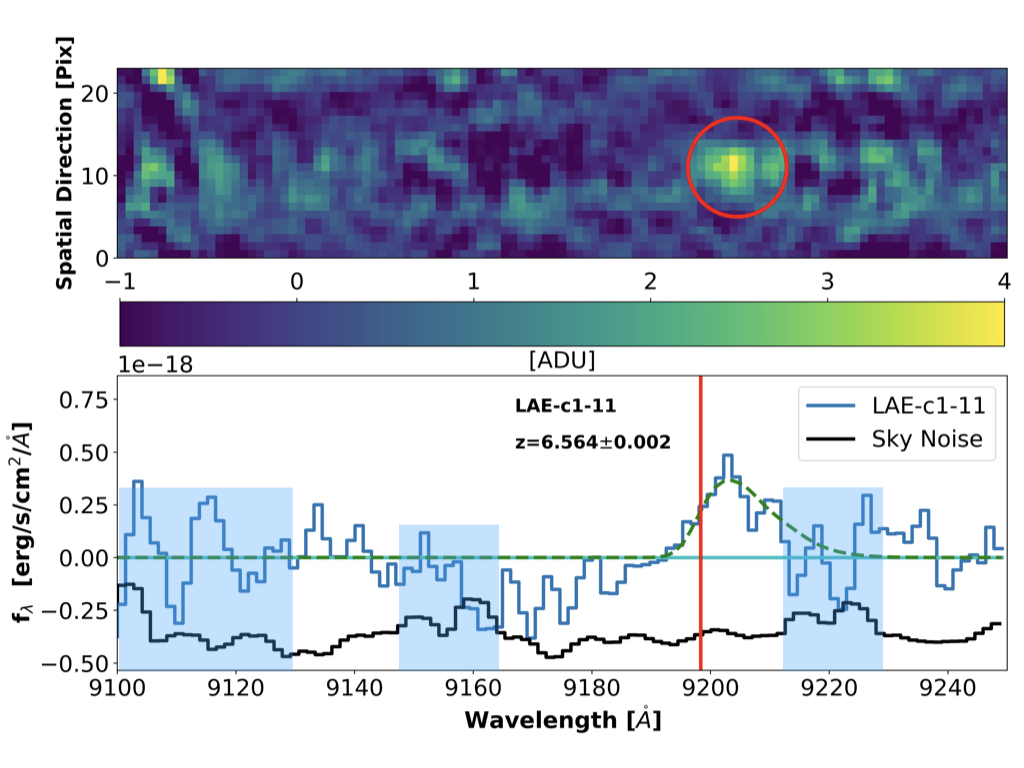}}
\subfigure[C2-20]{
\includegraphics[width=0.4\linewidth]{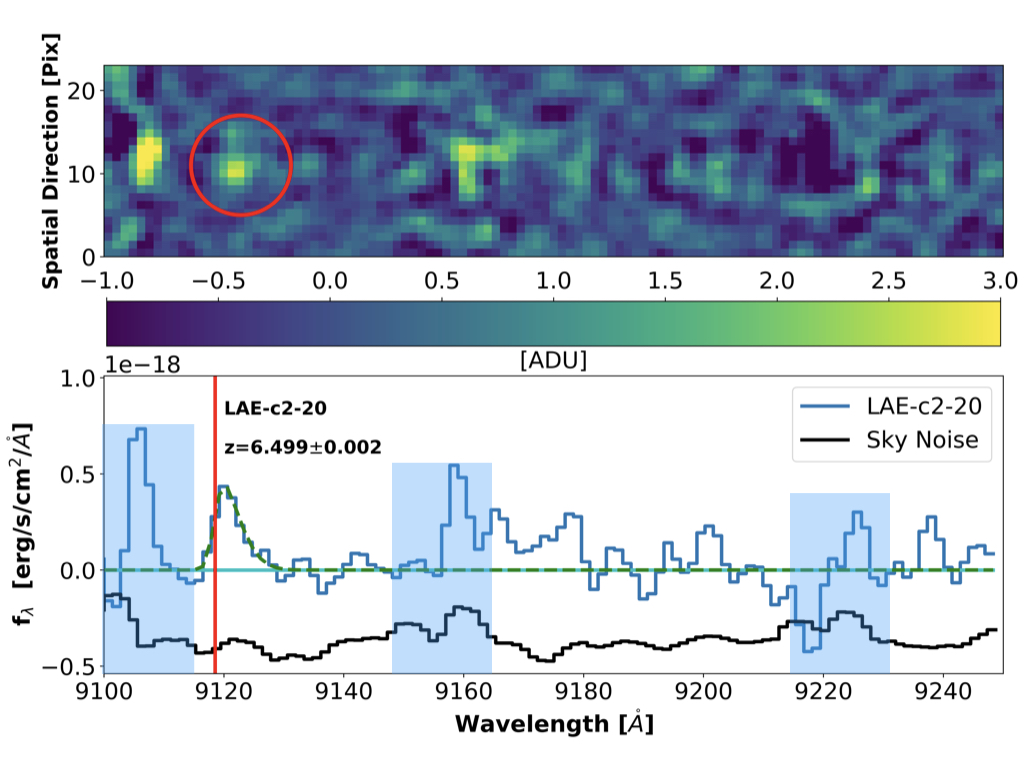}}
\subfigure[C2-35]{
\includegraphics[width=0.4\linewidth]{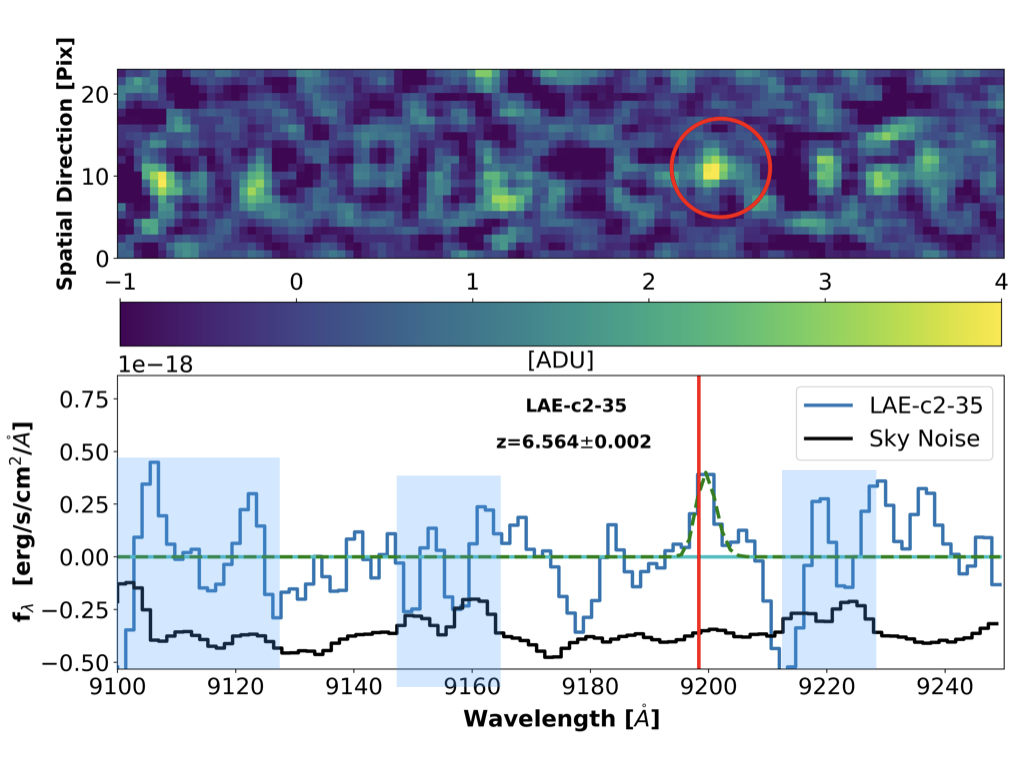}}
\subfigure[C2-43]{
\includegraphics[width=0.4\linewidth]{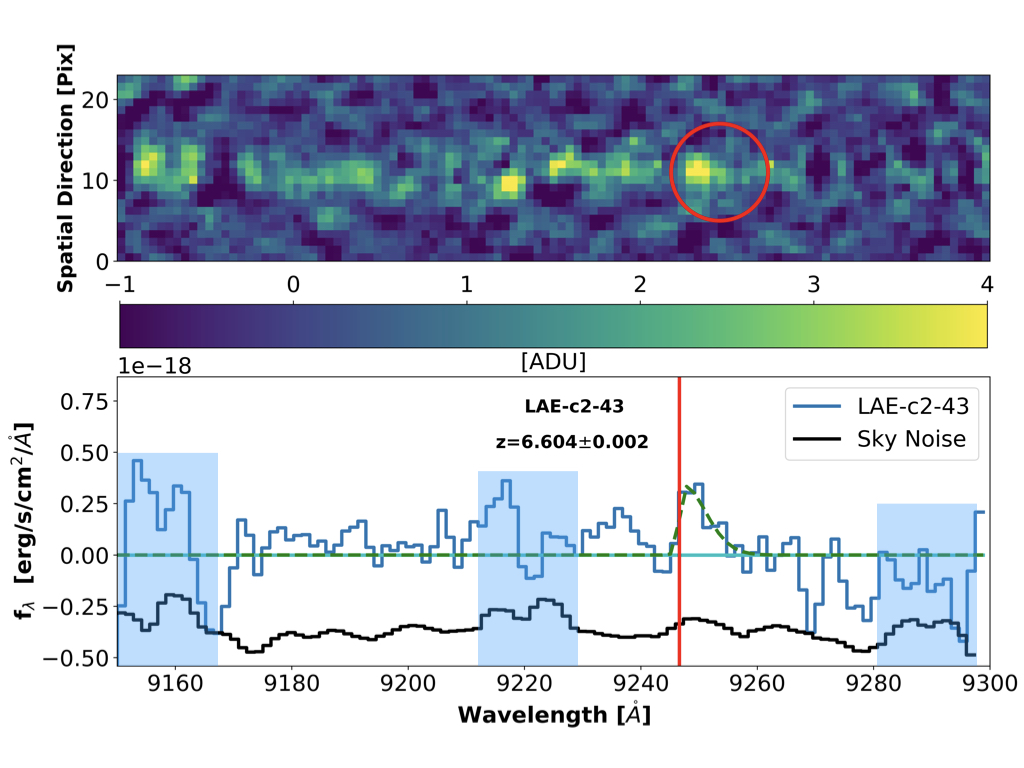}}
\subfigure[C2-46]{
\includegraphics[width=0.4\linewidth]{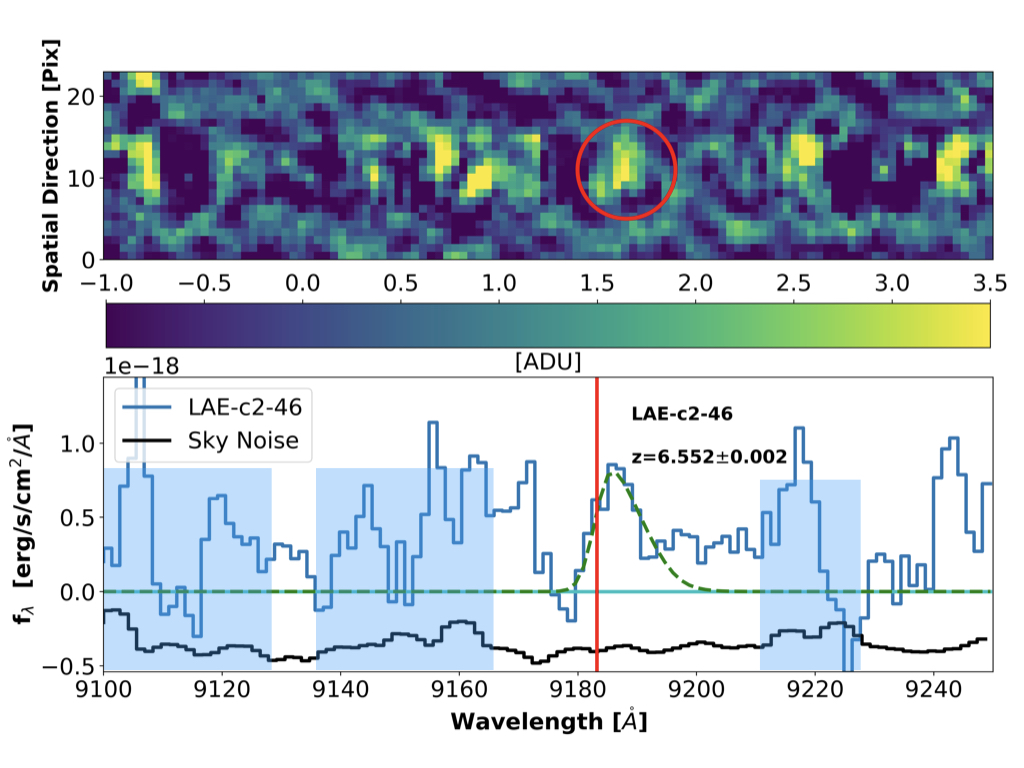}}
\caption{2D and 1D spectra of spectroscopically confirmed LAEs with the letter grade B. The blue solid lines are the extracted 1D spectra. The vertical red solid line indicates the spectroscopic redshifts of each object. The black solid lines are the sky background, measured using the same extraction window as the Ly$\alpha$ emission features, but scaled down to avoid interfering with the object's spectrum. The shaded blue areas indicate those regions where the sky lines are prominent.}
\label{gradeB}
\end{figure*}

\begin{figure*}
\centering
\subfigure[C1-06]{
\includegraphics[width=0.4\linewidth]{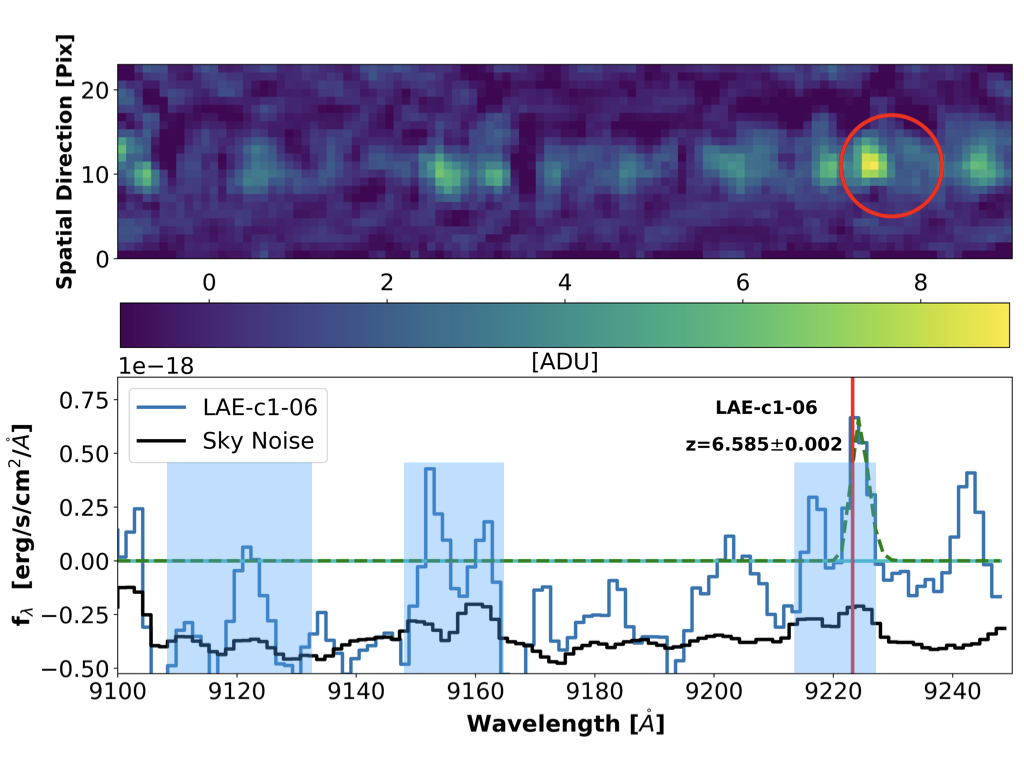}}
\subfigure[C1-07]{
  \includegraphics[width=0.4\linewidth]{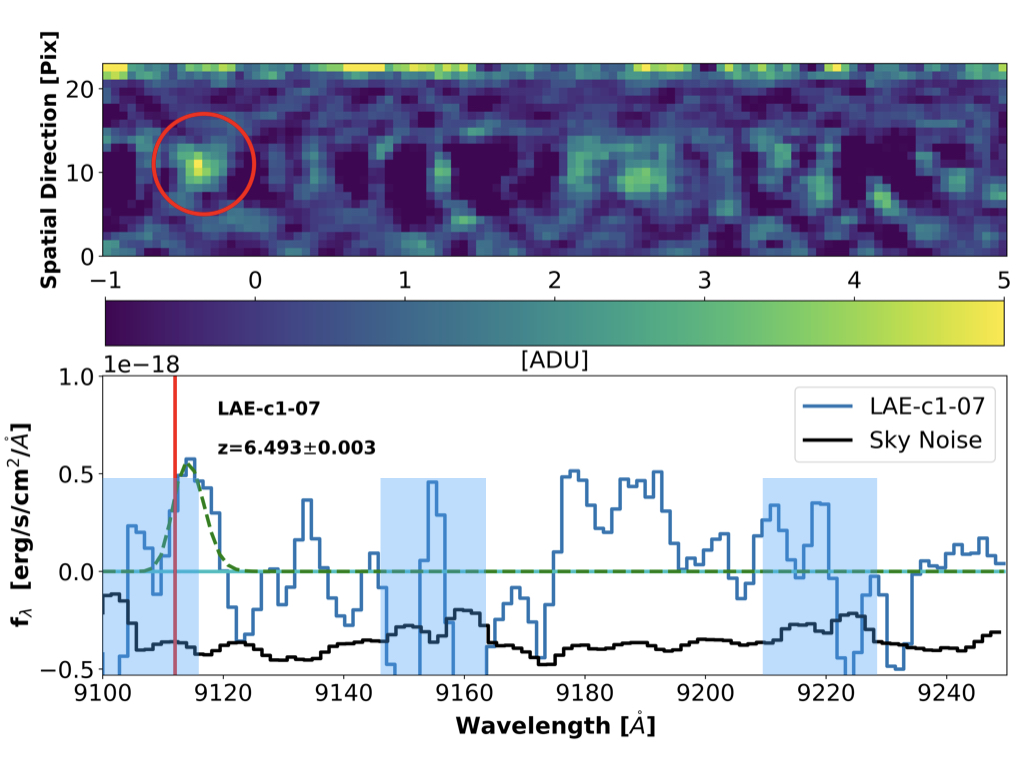}}
\subfigure[C2-17]{
\includegraphics[width=0.4\linewidth]{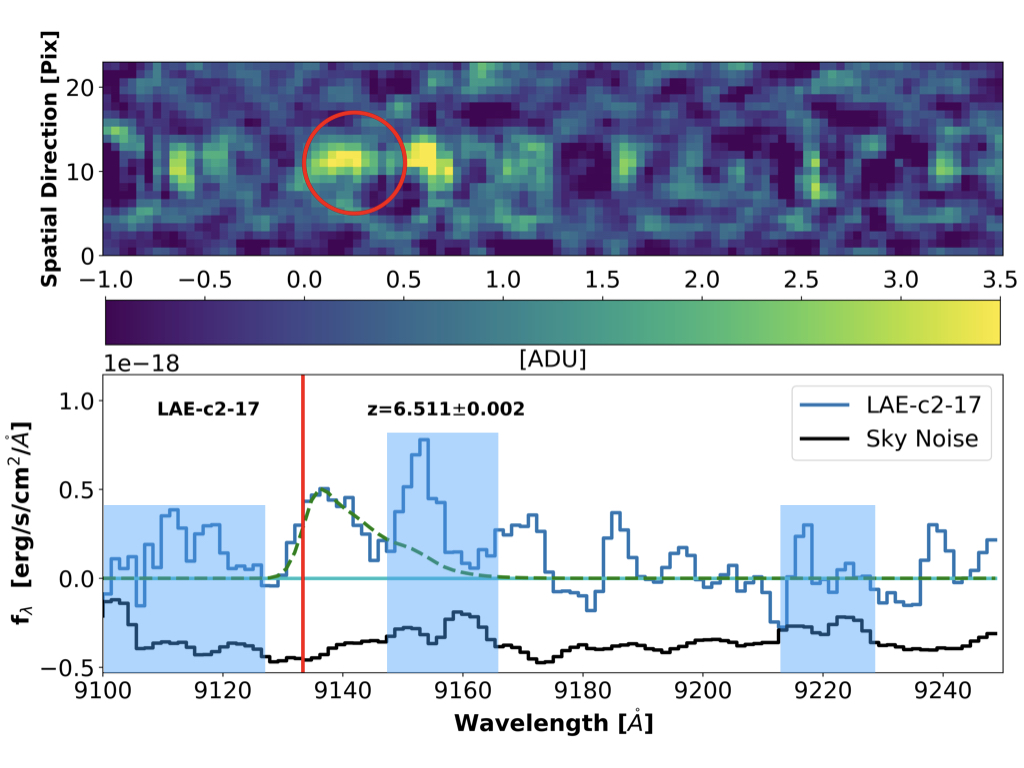}}
\subfigure[C2-26]{
\includegraphics[width=0.4\linewidth]{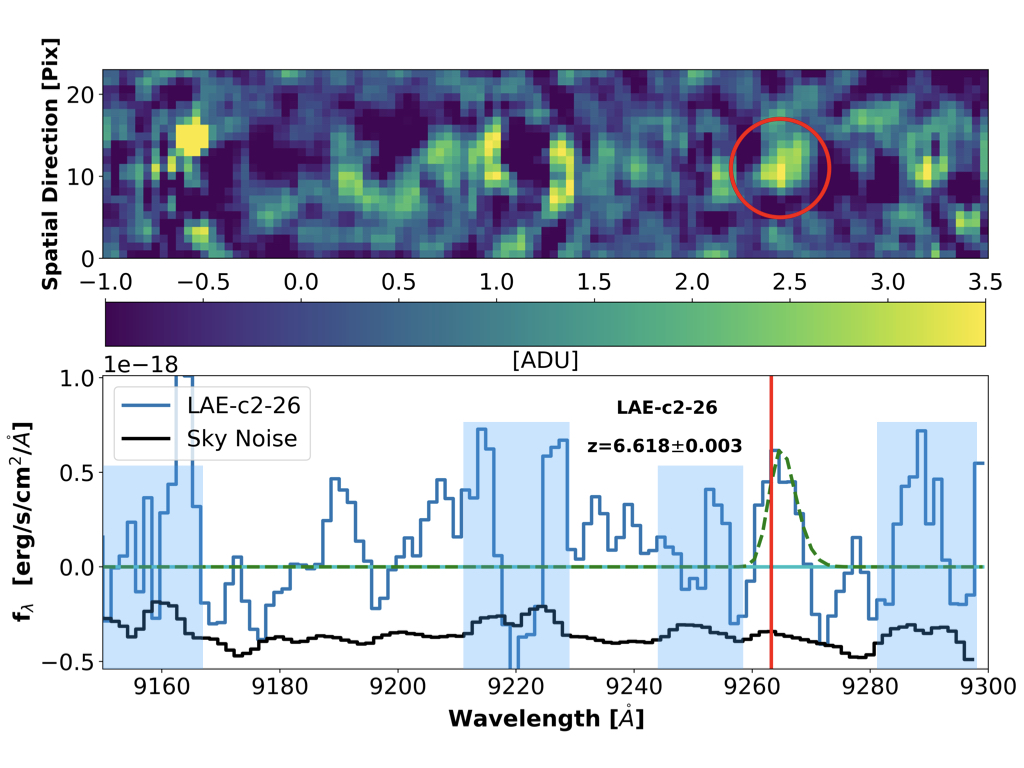}}
\caption{2D and 1D spectra of the LAE candidates with the letter grade C. The blue solid lines are the extracted 1D spectra. The vertical red solid line indicates the spectroscopic redshifts of each object. The black solid lines are the sky background, measured using the same extraction window as the Ly$\alpha$ emission features, but scaled down to avoid interfering with the object's spectrum. The shaded blue areas indicate those regions where the sky lines are prominent.}
\label{gradeC}
\end{figure*}

A co-added (stacked) spectrum was done by shifting all ten confirmed spectra to the same lambda-position ($\rm \lambda = 9130 \mathring{A}$) using their peaks as reference points  and performing $3\sigma$ clipping average of all frames. The stacked spectrum is  shown in Figure \ref{stack}. It exhibits enhanced S/N as compared to the individual galaxy spectra.  The stacked image shows an asymmetric emission line, with skewness factor 2.93$\pm 1.39$, while the noise of the continuum has substantially decreased, essentially because the noise and residuals from the sky line subtraction when averaged tend to zero. We compare the mean Ly$\alpha$ luminosity obtained from averaging the Ly$\alpha$ luminosities of all grade A and B LAEs ($\rm L^{mean}_{Ly\alpha}$) with the extracted Ly$\alpha$ luminosity of the stacked LAE ($\rm L^{stacked}_{Ly\alpha}$). Both values agree within the noise,  the $\rm L^{mean}_{Ly\alpha}$ is $\rm 2.52 \pm 1.33 \times 10^{42} ergs\:s^{-1}$, while the value of $\rm L^{stacked}_{Ly\alpha}$ is $\rm 2.00 \pm 0.50 \times 10^{42} ergs\:s^{-1}$. Since both $\rm L^{mean}_{Ly\alpha}$ and $\rm L^{stacked}_{Ly\alpha}$ are in good agreement within $\rm 1\sigma$ uncertainty, we are confident that we are not measuring anything other than the Ly${_\alpha}$ lines. Indeed, the asymmetry of the stacked line is noticeable. If sky lines were contaminating the stacked spectrum we would expect a more symmetric line. Which is not the case for our grade A and B LAEs. Therefore, we have high confidence in the 10 (both grade A and B) spectroscopically confirmed LAEs that we have detected.

We also compared the photometric and spectroscopic magnitudes in the filter F913 for our 10 confirmed LAEs (Figure \ref{difference}). We find that while for the grade A LAEs there is little difference between the two magnitudes, for the grade B LAEs the differences are more relevant. The sources that differ most are Ouchi's source C1-01 and C1-05.

\begin{table*}

\begin{tabular}{@{}l c c c c c c c}                       % 6 columns, alignment for each
        \hline
        $\:\:\:\:\:\:$Object   &   $EW_{0,\:Ly\alpha}$  &  $\rm f_{esc, \:Ly\alpha}$   &  SFR$\rm ^{phot}$      &   SFR$\rm ^{spec}$ & L$_{Ly\alpha,int}$  & L$_{1500}$ & M$_{1500}$   \\
        & [$\mathring{A}$]   &  & [$\rm M_{\odot} \:yr^{-1}$] & [$\rm M_{\odot}\: yr^{-1}$]  &  [$\rm 10^{42}\:ergs \:s^{-1}$] & [$\rm 10^{42}\:ergs \:s^{-1}\mathring{A}{-1}]$ &\\
      (1)  &  (2) &  (3) & (4) & (5) & (6) & (7) & (8)       \\
        \hline           
LAE-C1-01* &    	$>$83  & 	 0.19$\pm$0.10	&    53.7$\pm$27.9  &  13.7$\pm$0.5  &     78.9$\pm$ 41.6  & 1.36	 & -21.69	     \\ %A
LAE-C1-05 &     	$>$7    & 	 0.20$\pm$0.09	&    35.1$\pm$19.7  &  1.1$\pm$0.3    &     6.0$\pm$3.1       & 0.10	 & -18.95 	     \\ %B
LAE-C1-11 &    	$>$16  & 	0.18$\pm$0.09	&    15.2$\pm$8.2    &  2.6$\pm$0.4    &	 15.6$\pm$8.1     & 0.27	 & -19.87     	      \\ %B
LAE-C1-13 &    	$>$28  & 	0.20$\pm$0.10	&    16.6$\pm$9.2    &  4.7$\pm$0.5    &	 25.5$\pm$13.0   & 0.44	& -20.52     	      \\ %A
LAE-C1-15 &   	$>$13  & 	0.19$\pm$0.07	&    23.3$\pm$14.3  &  2.1$\pm$0.3    & 	 12.1$\pm$4.7     & 0.21	& -19.66    	      \\ %A
LAE-C2-20 &    	$>$8    & 	0.18$\pm$0.10	&      3.3$\pm$12.5  &   1.3$\pm$0.2   &     7.8$\pm$4.5       & 0.13	 & -19.12	       \\ %B
LAE-C2-29 &     	$>$18  & 	0.18$\pm$0.08	&    22.3$\pm$13.3  &  3.0$\pm$0.2    &	 18.3$\pm$15.2   & 0.32	& -20.05     	        \\ %A
LAE-C2-35  &    	$>$9    & 	0.26$\pm$0.14	&  126.3$\pm$65.5  &   1.5$\pm$0.7   &	 6.2$\pm$4.5       & 0.11	 & -19.26    	   	\\ %B
LAE-C2-40  &    	$>$23  & 	0.19$\pm$0.07	&    18.0$\pm$11.0  &  3.7$\pm$0.4    &	 21.6$\pm$8.2     & 0.37	 & -20.28      	   	\\ %A
LAE-C2-43  &   	$>$6    & 	0.19$\pm$0.08 	&    22.8$\pm$13.3  &   0.9$\pm$0.2   &	 5.3$\pm$2.5       & 0.09	 & -18.75   		\\ %B
LAE-C2-46  &   	$>$13  & 	0.20$\pm$0.07	&    18.5$\pm$11.2  &  2.2$\pm$0.6     &	 12.0$\pm$5.5     & 0.21	 & -19.70       	  \\ %B
	\hline
\end{tabular}
\caption{The list of all LAE candidates with their parameters; Columns: (1) name; (2) lower limit of the Ly$\alpha$ rest-frame equivalent width; (3) Ly$\alpha$ escape fraction;(4) star-formation rate estimated from $\rm L_{Ly\alpha}^{phot}$; (5) star-formation rate estimated from $\rm L_{Ly\alpha}^{spec}$; (6) Intrinsic Ly$\alpha$ Luminosity; (7) Luminosity of the continuum at 1500$\mathring{A}$; (8) Absolute magnitude at 1500$\mathring{A}$.}
\label{tab_sfr}
\end{table*}

\section{Discussion}

\subsection{Ly$_{\alpha}$ Equivalent Widths and Escape Fractions}

Deriving the Ly$\alpha$ rest-frame equivalent widths of the LAEs could be just a simple task, if any continuum flux would have been detected. However, in our case, the UV continuum was not detected, beyond $2\sigma$ level, for any of the LAEs, not even in the photometric observations, except for 3 objects \citep{Kritt17}. Thus, the only sound solution to estimate the rest-frame equivalent widths of the LAEs is to give lower limit values. The lower limit of the Ly$\alpha$ rest-frame equivalent width can be calculated as shown in Equation~\ref{ew_limit},

\begin{equation}
     EW^{lower}_{0,\:Ly\alpha}  = \frac{1}{1+z}\times \frac{F_{Ly\alpha}}{f^{2\sigma}_{cont}},
     \label{ew_limit}
\end{equation}

where $\rm EW^{lower}_{0,\:Ly\alpha}$, $\rm z$, $\rm F_{Ly\alpha}$, and $\rm f^{2\sigma}_{cont}$ are the lower limit of Ly$\alpha$ rest-frame equivalent width, redshift of a LAE, total Ly$\alpha$ emission flux, and the 2$\rm \sigma$ non-detected flux density level, respectively. We measured the value of $\rm f^{2\sigma}_{cont}$ by averaging $2\sigma$ RMS noise levels between 50 and 100 $\mathring{A}$ behind the Ly$\alpha$ emission lines on the 1D spectra of all 10 newly confirmed LAEs, which yield the value of $\rm 2.5\times10^{-19}\:ergs\:s^{-1}\:cm^{-2}\: \mathring{A}^{-1}$. The derived lower limit values of Ly$\alpha$ rest-frame equivalent widths for all LAEs observed spectroscopically are listed in Table~\ref{tab_sfr}. 

Note that \citet{Inoue18}, as part of the SILVERRUSH program, have demonstrated in a simulation that  hosting halo masses are closely related to Ly$\alpha$ production and the amount of neutral hydrogen gas. They compute Ly$\alpha$ escape fractions of LAEs at $\rm z=6.6$ that yield an excellent agreement with the observations of the SUBARU/HSC survey \citep{Konno18,Inoue18,Higuchi18}. We have thus adopted these simulations to estimate the f$_{esc,Ly_{\alpha}}$ based on the observed Ly$\alpha$ luminosities and an a priori LAE halo mass distribution (a detailed calculation is presented in \citet{Chanchaiworawit2019}. The Ly$\alpha$ escape fractions of the confirmed LAEs and LAE candidates estimated from their  Ly$\alpha$ luminosities are listed in Table~\ref{tab_sfr}. The average Ly$\alpha$ escape fraction of the 10 LAEs is $0.19\pm0.09$, which is in good agreement with the literature (e.g., \citet{Ouchi10,Inoue18}). \citet{Hayes2010}, for example, predict an average Ly$\alpha$ escape fraction at $\rm z=6.6$ of around 0.2.

\subsection{Ly$\alpha$ Star Formation Rates} \label{sec:sfr}

 To compute the star formation rates (SFR) we have followed two different approaches. First we have assumed the standard Kennicutt calibration \citep{Kennicut1998}, assuming Case B recombination and a Salpeter Initial Mass Function (IMF) with a mass range of 0.1 to 100 $\rm M_{\odot}$, considering a $\frac{F_{Ly\alpha}}{F_{H\alpha}}$ flux ratio of 8.7. This method follows a traditional and straightforward way of computing star formation rates, but it should be noted that it is based on the assumption of a star formation episode producing stars at a constant rate during tens of Myr, until the birth and death of the most massive ionising stars reach an equilibrium. We first derived the star formation rate for each LAE directly from its observed Ly$_\alpha$ luminosity.  No attempts were made to correct for internal extinction (within the LAEs) as we do not have ways of determining the extinction. Besides, the expected extinction of LAEs at z$\sim 6.5$ should not be high. Note that the SFRs thus obtained are very sensitive to the assumption of the IMF and the range of masses used. For instance, \citet{Oti2010} have shown that changing the mass ranges through the variety of choices available in the literature could vary the Ly$\alpha$ luminosity by a factor of 4, implying that the star formation rates derived from the Ly$\alpha$ luminosities need to be related to a given specific IMF and mass range.

The stars producing the Ly$\alpha$ flux and continuum do so before any scattering or absorption processes within the neutral medium. Therefore, the intrinsic Ly$\alpha$ luminosity is different from the observed Ly$\alpha$ luminosity and could be retrieved by dividing the observed Ly$\alpha$ luminosity by the Ly$\alpha$ escape fraction. Likewise, we can derive intrinsic Star Formation Rates (SFR) using Equation~\ref{eq:SFR},

\begin{equation}
\centering
 \frac{SFR_{Ly\alpha}}{M_\odot\:yr^{-1}} = 9.1\times10^{-43}\:\frac{L_{Ly\alpha}^{obs}}{ergs \:s^{-1}} \times \frac{1}{f_{esc,Ly\alpha}},
\label{eq:SFR}
\end{equation}
where $\rm L_{Ly\alpha}^{obs}$ is the observed Ly$\alpha$ luminosity, and $\rm f_{esc,Ly\alpha}$ is the corresponding Ly$\alpha$ escape fraction. The intrinsic Ly$\alpha$ luminosities and SFRs of LAEs are shown in Table~\ref{tab_sfr}. 

The second approach assumed the current star formation episode to be essentially coeval. In this scenario we can estimate the star formation strength (SFS), which is the mass of gas converted into stars \citep{Oti2010} in the current burst. Deriving the average value of the intrinsic Ly$_{\alpha}$ luminosity, for the 10 sources with good grade, we get $\rm \sim 1.30\times 10^{43} ergs\: s^{-1}$. We can then use the calibration by \citep{Oti2010} (available also as a Webtool\footnote{http://www.laeff.cab.inta-csic.es/research/sfr/}), to get the average mass ($\rm 2.52\times 10^8 \: M_{\odot}$) that has been converted into stars in the current burst of star formation (assuming a coeval starburst at 4 Myr, with a Salpeter IMF from 0.1 -- 100 M$_{\odot}$). If we assume that these galaxies have an average  mass of around 10$^9\: M_{\odot}$, about 25\% of all of the gas in these galaxies has been converted into stars in the current burst in a very short time, which is a significant fraction of their total mass.  

While the derivation of star formation rates based on calibrations which assume a long star formation episode in equilibrium is usual in the literature, this might not reflect properly the conditions of intense starburst in high redshift galaxies. For Instance, \citet{Oti2010} have shown that once this equilibrium phase has been reached, the intrinsic  Ly$_{\alpha}$ equivalent width can never be above around 100\AA\, while \citet{Sobral2018, harikane2018} have shown, analysing a large sample of galaxies with measured values of the escape fraction, that their intrinsic equivalent widths peak at around 200\AA. Indeed, these values are only consistent with very young, almost coeval star formation episodes.

\subsection{1500\AA\ Continuum Luminosities}

The rest-frame Ultra-Violet (UV) continuum of galaxies, when detected, is commonly used to provide an additional estimate of the star formation rate. Most of the energy radiated from young, massive stars is emitted in the UV range (from less than 912 to around 3000 $\mathring{A}$). While the UV continuum is potentially affected by dust extinction, an effect which should be negligible at high redshift (see for example \citet{Hayes2010}), it is not affected at all by scattering or absorption processes in the neutral gas, as the Lyman$\alpha$ emission is. Thus, the UV luminosity serves as a reliable estimator of the overall star formation rate. However, the LAEs we are dealing with are quite faint, and the observed spectra were not deep enough to recover their rest-frame UV continuum.

Just for reference we have computed expected continuum luminosities at 1500\AA\ for the A and B-grade sources, using the models by \citet{Oti2010}. These calibrations are based on a self--consistent evaluation of the most used star formation rate tracers at different wavelengths, and for different star formation scenarios. Based on these models, the expected value of any other star formation rate tracer can be derived based on the observed value of any other. For a given value of the star formation parameters such as age, star formation regime, IMF, etc. ... Following this approach we have computed the expected luminosities at 1500\AA\ as constrained by the intrinsic Ly$_\alpha$ luminosities. As long as the Ly$_\alpha$ escape fractions we have used are reliable, they would have taken into account all effects related with Ly$_\alpha$ photons scattering and absorption by the neutral interstellar medium, and should be a valid prior to provide a first estimate of the continuum luminosities. Following the above discussion, they have been computed for a coeval starburst of 4~Myr age. The estimated $\rm 1500\mathring{A}$ continuum luminosities are listed in Table \ref{tab_sfr}. We also list the $1500\mathring{A}$ absolute magnitudes, which with an averaged M$_{1500} = -19.6$ these are indeed sub L* galaxies \citep{Bouwens2006,Ouchi2009}.

\subsection{Mechanical Energy}
Studying the mechanical energy released  by star formation episodes in the galaxies in the proto-cluster is interesting as this mechanical energy might be punching holes in the surrounding neutral gas. Indeed, this is required to produce the relatively large Ly$\alpha$ escape fractions that we have derived. \citet{Oti2010} computed the mechanical energy yielded by stellar winds and supernovae. A small fraction of this mechanical energy ends up heating the gas by collision and is transformed into X-rays. With the help of the above mentioned Webtool, we could determine the average X-ray luminosity, corresponding to the average value of the intrinsic Ly$\alpha$ luminosity of the 10 good grade galaxies. This is assuming a coeval starburst of around 4~Myr age. The average X-ray luminosity for the 10 good grade sources is $\rm L_X (0.4 - 2.4 KeV) = 3.69\times 10^{41}\: ergs\:s^{-1}$. Typically, about 5\% of the mechanical energy yielded by stellar winds and supernovae is transformed into X-ray emission.  Thus, the estimated mechanical energy input results in $\rm 7.4\times 10^{42}\: ergs\:s^{-1} $, which is significantly larger than the mechanical energy produced by low redshift, local starburst galaxies. For instance, \citet{Silich2009} give a value for the mechanical energy produced by a starburst galaxy, such as M82, of $\rm 2.5 \times 10^{40}\: ergs\:s^{-1}$. This is two orders of magnitude lower.  Therefore, we conclude that the typical LAEs in this group release huge quantities of mechanical energy, sufficient to pierce holes in the circumgalactic medium. That allows the ionising photons from these starburst galaxies to escape into the intergalactic medium (IGM), eventually ionising a large region outside the circumgalactic medium.

\vspace{0.1in}

\section{Conclusions}
We have spectroscopically confirmed 10 LAEs from the SXDX-Newton field, which were previously detected by us, photometrically. From a  GTC/OSIRIS mask with 17 science slits devoted to LAE candidates, we obtained a spectroscopic success rate of $\rm \sim \frac{2}{3}$, consistent with the previously determined spuriousness value \citep{Kritt17}. Nevertheless, the remaining LAE candidates were also detected with lower reliability level. 

We have measured their Ly$_{\alpha}$ fluxes and associated star formation rates (or star formation strengths for coeval bursts). The intrinsically intermediate to high SFRs have clearly demonstrated the signatures of high redshift, young starburst galaxies.  Moreover, the lower limits of Ly$\alpha$ equivalent widths are relatively high, as expected from sources that are undergoing young starbursts. We have computed values for the intrinsic Ly$\alpha$ luminosities, and from these we have determined the expected continuum luminosities and absolute magnitudes, showing that indeed the detected sources are low luminosity sources. Finally, we have derived the mechanical energy that these sources are outputting onto the CGM, finding that it is substantially high, which suggests the formation of holes, that allow ionising radiation to escape. The escape of ionising photons from these young starburst galaxies may be an important driver of the Re-ionisation of the Universe. 

\bigskip
Acknowledgements

RC \& JMRE acknowledge support from the AEI of the Spanish Ministry of Science under grant AYA2015-70498-C2-1, and AYA2017-84061-P (both co-founded with FEDER funds.) We thank the GTC support staff for performing the observations. We acknowledge very valuable comments by an anonymous referee that allowed a great improvement of the manuscript. KC recognises the Royal Thai Government Scholarship in the section of the Ministry of Sciences \& Technologies for the National Astronomical Research Institute of Thailand (NARIT) and the University of Florida's Graduate School Doctoral Dissertation Award for guidance and financial support to conduct this work and complete his thesis. ESS \& AMO acknowledge funding and support from the Spanish MINECO under the projects MDM-2014-0369 of ICCUB (Unidad de Excelencia `Mar\'{i}a de Maeztu'), AYA2015-70498-C02-2-R and AYA2017-88085-R (both co-funded with FEDER funds), and the Catalan DEC grant 2017SGR643. JMRE also acknowledges support from the Spanish Ministry of Science through AYA2017-84061-P grant. AH, RC, \& JMRE acknowledge the Spanish Ministry of Science's support through the Severo Ochoa grant SEV2015-0548. AH acknowledges support by the Spanish Ministry of Science through grant AYA2015-68012-C2-1 and by the Gobierno de Canarias through grant ProID2017010115. JG acknowledges support by the Spanish Ministry of Science through grant AYA2016-75808-R. JMMH is funded by Spanish State Research Agency grants ESP2017-87676-C5-1-R and MDM-2017-0737 (Unidad de Excelencia Mar\'{\i}a de Maeztu CAB). This work was partly done using GNU Astronomy Utilities (Gnuastro, ascl.net/1801.009) version 0.3. Work on Gnuastro has been funded by a Japanese Ministry of Education, Culture, Sports, Science, and Technology (MEXT) scholarship and its Grant-in-Aid for Scientific Research (21244012, 24253003), and the European Research Council (ERC) advanced grant 339659-MUSICOS.

\bibliographystyle{mnras}
\bibliography{bibrosa2}

\label{lastpage}

\end{document}